\documentclass{aa}
\usepackage[varg]{txfonts}
\usepackage{mathptmx}
\usepackage{adjustbox,lipsum}

\usepackage{graphicx}
\usepackage{amstext}

\begin{document}

\title{The spectral energy distribution of galaxies at $z > 2.5$:
  Implications from the {\it H\lowercase{erschel}}/SPIRE color-color diagram}

\author{Yuan Fangting\inst{1}, Veronique Buat\inst{2}, Denis Burgarella\inst{2}, Laure Ciesla\inst{3,4},
  Sebastien Heinis\inst{2,5}, Shiyin Shen\inst{1},
  Zhengyi Shao\inst{1}, Jinliang Hou\inst{1}}

\institute{Key Laboratory for Research in Galaxies and
  Cosmology, Shanghai Astronomical Observatory, CAS, 80 Nandan Road,
  Shanghai 200030, China 
  \email{yuanft@shao.ac.cn}
  \and
  Aix Marseille Universit\'{e}, CNRS, LAM (Laboratoire
  d'Astrophysique de Marseille) UMR 7326, 13388, Marseille, France
  \and
  University of Crete, Department of Physics, Heraklion
  71003, Greece
  \and
  Institute for Astronomy, Astrophysics, Space Applications and Remote Sensing, National Observatory of Athens, GR-15236 Penteli, Greece
  \and
  Department of Astronomy, University of Maryland,
  College Park, MD20742-2421, USA}

\date{Received,2014; accepted,2014}

\abstract
{We use the {\sl Herschel} SPIRE color-color diagram to study the
spectral energy distribution (SED) and the redshift estimation of high-z
galaxies. We compiled a sample of 57 galaxies with spectroscopically
confirmed redshifts and SPIRE detections in all three bands at $z=2.5-6.4$, and compared their average SPIRE colors with SED templates
from local and high-z libraries. We find that local SEDs are
inconsistent with high-z observations. The local calibrations of the
parameters need to be adjusted to describe the average colors of high-z 
galaxies. For high-z libraries, the templates with an evolution from z=0 
to 3 can well describe the average colors
of the observations at high redshift. Using these templates, we defined 
color cuts to divide the SPIRE
color-color diagram into different regions with different mean
redshifts. We tested this method and two other color cut methods using a
large sample of 783 Herschel-selected galaxies, and find that although
these methods can separate the sample into populations with
different mean redshifts, the dispersion of redshifts in each population is
considerably large. Additional information is needed for better sampling. 
}

   \titlerunning{The SED of galaxies at $z > 2.5$}
   \authorrunning{Yuan et al. (2014)}
   \maketitle
%

\section{Introduction}
\label{sec:intro}
The {\it Herschel} Space Observatory, with its 3.5 meter mirror and
the two  PACS \citep{poglitsch2010} and SPIRE
\citep{griffin2010} instruments, covers the electromagnetic spectrum
from 55 to 672 $\mu$m. \textit{Herschel} provides us with an unprecedented
chance to study the infrared (IR) emission of galaxies. The SPIRE
instrument is able to  detect galaxies up to quite high redshift
($z\sim3$ to $z\sim7$), where the $250$, $350,$ and $500$ $\mu$m bands
of SPIRE corresponds to the peak of the IR spectral energy distribution (SED)
of star-forming galaxies. The combination of the three band fluxes provide
constraints on the shape of IR SEDs of these high-z objects and on physical 
parameters like the dust temperature or the dust mass
\citep[e.g.,][]{amblard2010,chapman2010,magnelli2010,roseboom2012}.

The identification of IR sources and the measure of their redshift
are challenging at $z>2.5$. The obscuration of these high-z IR
luminous galaxies makes their optical
emission very faint and hard to detect. The large beam size of SPIRE
also makes the cross-matching with optical sources very difficult.
As a consequence, acquiring
spectroscopic or photometric redshifts from optical bands is not
possible for most of these sources. Submillimeter spectroscopic
redshifts can be used for
these high-z IR sources: the redshifts of most spectroscopically
confirmed high-z sources are measured in this way
\citep[e.g.,][]{riechers2013,dowell2014}. However, the
method is not yet efficient enough to obtain redshifts for large galaxy
samples \citep{casey2014}. Consequently, it is certainly useful to directly
estimate redshifts for large samples of galaxies only detected in IR.

The IR data from $Herschel$ alone are not sufficient to provide reliable
photometric redshifts for individual galaxies, but they can be useful in providing
redshift distributions of large
samples and candidates for follow-up spectroscopic
observations.  Several works have been dedicated
to presenting methods to select high-z candidates
using the SPIRE data \citep[e.g.,][]{amblard2010,pope2010,roseboom2012}. Based on the redshift
distributions of IR galaxies on a SPIRE color-color diagram,
a color cut is  applied to select high-z and low-z galaxy populations
\citep[e.g.,][]{amblard2010,pope2010,riechers2013,dowell2014}. All these
works are based on modified blackbody (MBB) models to model the IR emission.
It is necessary to reinvestigate this topic
with updated libraries of more sophisticated IR SED templates, which are now available using {\it Herschel} observations.

Empirical template of IR SEDs ($8-1000\mu$m)
are  widely used to derive the total IR luminosity, the SFR, and the dust properties of galaxies. Prior to {\sl Herschel}, local
templates were built based on IRAS, ISO, SCUBA, or {\sl Spitzer}
\citep[e.g.,][]{ce2001,dh2002,dl2007,rieke2009}. Adding
  {\sl Herschel} data, recent studies have constructed new IR
templates for low- and high-redshift objects
\citep{elbaz2011,smith2012,magdis2012,ciesla2014}.
However, there are still no IR SED templates for galaxies at $z>3$.
Numerous studies still use local templates or MBB
models when deriving the properties of very distant galaxies
\citep[e.g.,][]{cox2011,conley2011,combes2012,huang2014,robson2014}.

To obtain a clearer and up-to-date view of where we stand in this post-{\it Herschel} era,
we compare the SED templates
from different libraries with a sample of high-z galaxies that have
spectroscopic redshifts and reliable SPIRE flux measurements. We
attempt to find the SED templates that can best describe the average
SPIRE colors of high-z galaxies. Through a comparison with the local
calibrations, we can test the evolution of the galaxy populations. In
addition, we examine whether different redshift populations could be
separated by using a SPIRE color-color diagram.

The paper is organized as follows. We first introduce the samples and
the SED libraries used in this study in Sections \ref{sec:data} and
\ref{sec:sedtemp}. We compare SED templates with observed
high-z data on a color-color diagram at different redshifts in
Section \ref{sec:res}. In Section \ref{sec:discussion} we discuss the
possible influence of the detection limits of SPIRE observations
and the ability of applying SPIRE color cuts to select high-z
galaxies. The main conclusions are summarized in Section
\ref{sec:conc}.

Throughout this paper, $S_{\lambda}$ [mJy] means the flux density
$S_{\nu}$ at $\lambda$
$\mu$m. The IR luminosity $L_{\rm IR}$ [$L_{\odot}$] is integrated
from $8$ to $1000$ $\mu$m of the spectrum. The $\Lambda${--}cosmology
is adopted: $\Omega_{m}$=0.3, $\Omega_{\Lambda}$=0.7, and $H_{0}$=$70$
km s$^{-1}$ Mpc$^{-1}$.

\section{Sample}
\label{sec:data}
\subsection{A spectroscopically confirmed sample of galaxies at $z>2.5$}

\begin{table*}
\caption{High-z galaxies at $z\ \geq\ 3$ from literature.}
\centering
\begin{tabular}{cccccccc}
\hline
Reference & Objects & z & $S_{250}$ [mJy] & $S_{350}$ [mJy] &
$S_{500}$ [mJy] & $\log(L_{\rm IR}[L_{\odot}])$\tablefootmark{a} & $\mu$\\
\hline
  Weiss+2013 & SPT0103-45 & 3.0917 & 121.0$\pm$15.0 & 210.0$\pm$23.0 & 222.0$\pm$24.0 & 13.84 & 1.0   \\
  Weiss+2013 & SPT0345-47 & 4.2958 & 242.0$\pm$25.0 & 279.0$\pm$29.0 & 215.0$\pm$23.0 & 14.3 & 1.0   \\
  Weiss+2013 & SPT0346-52 & 5.6559 & 136.0$\pm$16.0 & 202.0$\pm$22.0 & 194.0$\pm$21.0 & 14.36 & 5.4   \\
  Weiss+2013 & SPT0418-47 & 4.2248 & 115.0$\pm$14.0 & 189.0$\pm$20.0 & 187.0$\pm$20.0 & 14.08 & 21.0 \\
  Weiss+2013 & SPT0441-46 & 4.4771 & 62.0$\pm$10.0 & 98.0$\pm$12.0 & 105.0$\pm$13.0 & 13.87 & 1.0   \\
  Weiss+2013 & SPT0459-59 & 5.7993 & 35.0$\pm$10.0 & 54.0$\pm$10.0 & 61.0$\pm$11.0 & 13.83 & 1.0   \\
  Weiss+2013 & SPT0529-54 & 3.3689 & 74.0$\pm$13.0 & 137.0$\pm$17.0 & 162.0$\pm$19.0 & 13.75 & 9.4   \\
  Weiss+2013 & SPT0532-50 & 3.3988 & 214.0$\pm$23.0 & 269.0$\pm$28.0 & 256.0$\pm$27.0 & 14.08 & 1.0   \\
  Weiss+2013 & SPT2103-60 & 4.4357 & 43.0$\pm$10.0 & 72.0$\pm$11.0 & 108.0$\pm$15.0 & 13.81 & 1.0   \\
  Weiss+2013 & SPT2146-55 & 4.5672 & 58.0$\pm$12.0 & 79.0$\pm$14.0 & 82.0$\pm$14.0 & 13.8 & 1.0   \\
  Weiss+2013 & SPT2147-50 & 3.7602 & 73.0$\pm$12.0 & 114.0$\pm$14.0 &
  116.0$\pm$15.0 & 13.77 & 1.0   \\
  Weiss+2013 & SPT0113-46 & 4.2328  &  22$\pm$8 & 54$\pm$10 &
  82$\pm$11 & 13.63 & 1.0 \\
  Weiss+2013 & SPT0125-47 & 2.51480 &  785$\pm$79 & 722$\pm$73 &
  488$\pm$50 & 14.37 & 1.0 \\
  Weiss+2013 & SPT0243-49 & 5.699   &  18$\pm$8  & 26$\pm$8  &
  59$\pm$11 & 14.06 & 1.0 \\
  Weiss+2013 & SPT2132-58 & 4.7677  &  55$\pm$11 & 75$\pm$12 &
  78$\pm$12 & 13.81 & 1.0 \\
  Weiss+2013 & SPT2134-50 & 2.7799  &  346$\pm$36 & 339$\pm$35 &
  257$\pm$28 & 14.09 & 1.0 \\
  Magnelli+2012 & LESS011 & 2.679 & 15.2$\pm$2.6 & 19.2$\pm$3.4 & 17.9$\pm$4.2 & 12.73 & 1.0   \\
  Magnelli+2012 & LOCK850.14 & 2.611 & 19.4$\pm$3.8 & 24.9$\pm$5.7 & 30.6$\pm$8.3 & 12.82 & 1.0   \\
  Magnelli+2012 & AzLOCK.10 & 2.56 & 12.4$\pm$3.9 & 34.5$\pm$7.5 & 35.0$\pm$8.7 & 12.8 & 1.0   \\
  Magnelli+2012 & SMMJ14011$+$0252 & 2.565 & 61.7$\pm$6.0 & 63.1$\pm$3.8 & 48.5$\pm$4.0 & 13.27 & 3.0   \\
  Magnelli+2012 & SMMJ14009$+$0252 & 2.934 & 66.4$\pm$6.0 & 65.8$\pm$3.8 & 53.7$\pm$4.2 & 13.41 & 1.5   \\
  Magnelli+2012 & SMMJ00266$+$1708 & 2.73 & 52.8$\pm$7.2 & 61.3$\pm$4.5 & 44.6$\pm$4.7 & 13.28 & 2.4   \\
  Magnelli+2012 & SMMJ04542$-$0301 & 2.911 & 76.0$\pm$15.1 & 94.4$\pm$8.8 & 84.1$\pm$11.5 & 13.5 & 50.0 \\
  Magnelli+2012 & SMMJ16354$+$66114 & 3.188 & 20.5$\pm$6.0 & 25.7$\pm$3.0 & 17.4$\pm$3.3 & 13.03 & 1.7   \\
  Magnelli+2012 & SMMJ16355$+$66122B & 2.516 & 78.1$\pm$6.0 & 70.2$\pm$7.2 & 74.7$\pm$3.5 & 13.34 & 22.0 \\
  Magnelli+2012 & SMMJ16355$+$66123A & 2.516 & 53.7$\pm$6.0 & 47.1$\pm$3.1 & 26.9$\pm$3.5 & 13.23 & 14.0 \\
  Magnelli+2012 & SMMJ02399$-$0136 & 2.81 & 65.7$\pm$6.1 &
  72.0$\pm$3.9 & 62.8$\pm$4.6 & 13.38 & 2.5   \\
  Casey+2012 &  X24 J033136.96$-$275510.9 & 3.145 & 15.5$\pm$3.8 & 20.8$\pm$3.8 & 24.0$\pm$4.8 & 12.91 & 1.0   \\
  Casey+2012 &  X1.4 J033151.94$-$275326.9 & 2.938 & 12.9$\pm$3.9 & 21.6$\pm$3.7 & 24.2$\pm$4.2 & 12.83 & 1.0   \\
  Casey+2012 &  X24 J095916.08$+$021215.3 & 4.454 & 25.8$\pm$2.2 & 24.1$\pm$2.9 & 14.7$\pm$3.2 & 13.3 & 1.0   \\
  Casey+2012 &  X1.4 J100111.52$+$022841.3 & 3.975 & 24.5$\pm$2.2 & 32.9$\pm$4.3 & 22.8$\pm$6.8 & 13.27 & 1.0   \\
  Casey+2012 & X24 J100150.16$+$024017.2 & 2.883 & 13.8$\pm$2.2
  & 12.2$\pm$2.9 & 8.9$\pm$4.0 & 12.71 & 1.0 \\
  Casey+2012 & X1.4J100008.64$+$022043.1 & 2.888 & 16.5$\pm$2.2
  & 12.3$\pm$2.9 & 9.4$\pm$3.6 & 12.79 & 1.0 \\
  Saintonge+2013 & 8:00arc & 2.73 & 91.3$\pm$3.6 & 78.5$\pm$4.2 & 47.8$\pm$4.0 & 13.5 & 12.3 \\
  Saintonge+2013 & cB58 & 2.729 & 24.9$\pm$2.4 & 16.0$\pm$3.3 &
  10.9$\pm$3.6 & 12.97 & 31.8 \\
  Saintonge+2013 & J0712 & 2.646 & 9.1$\pm$2.9 &  3.8$\pm$3.4 &
  4.7$\pm$3.1 & 12.5 & 27.5 \\
  Saintonge+2013 & J1226 & 2.925 & 9.7$\pm$5.3 &  3.7$\pm$3.3 &
  3.1$\pm$5.1  & 12.49 & 40.0 \\
  Saintonge+2013 & Eye   & 3.074 & 26.8$\pm$7.0 & 33.1$\pm$8.7 &
  19.1$\pm$10.0 & 13.08 & 30.0 \\

  Dowell+2014 & LSW20 & 3.36 & 17.6$\pm$0.0 & 36.6$\pm$0.0 & 43.9$\pm$0.0 & 13.17 & 1.0   \\
  Dowell+2014 & FLS1 & 4.29 & 62.3$\pm$0.0 & 94.0$\pm$0.0 & 98.2$\pm$0.0 & 13.81 & 1.0   \\
  Dowell+2014 & FLS5 & 4.44 & 24.7$\pm$0.0 & 45.1$\pm$0.0 & 48.2$\pm$0.0 & 13.5 & 1.0   \\
  Dowell+2014 & LSW102 & 5.29 & 49.7$\pm$0.0 & 118.1$\pm$0.0 & 140.4$\pm$0.0 & 14.11 & 1.0   \\

  Roseboom+2012 & G1 & 4.05 & 18.6$\pm$2.3 & 31.7$\pm$2.7 & 24.3$\pm$2.7 & 13.23 & 1.0   \\
  Roseboom+2012 & G3 & 2.98 & 46.3$\pm$2.0 & 46.5$\pm$2.3 & 32.7$\pm$2.3 & 13.28 & 1.0   \\
  Roseboom+2012 & G19 & 2.91 & 11.2$\pm$2.1 & 11.3$\pm$2.4 &
  17.3$\pm$2.4 & 12.65 & 1.0   \\
  Roseboom+2012 &  L1 & 2.562 & 80.1$\pm$3.4 & 63.2$\pm$3.8 &
  40.6$\pm$4.4 & 13.4 & 1.0 \\
 Negrello+2010 & SDP81 & 3.04 & 129.0$\pm$20.0 & 182.0$\pm$28.0 & 166.0$\pm$27.0 & 13.79 & 19.0 \\
  Negrello+2010 & SDP130 & 2.63 & 105.0$\pm$17.0 & 128.0$\pm$20.0 & 108.0$\pm$18.0 & 13.54 & 6.0   \\
  Smolcic+2015 & AzTEC1 & 4.3415 & 19.5$\pm$6.0 & 29.8$\pm$7.6 & 28.8$\pm$9.0 & 13.31 & 1.0   \\
  Smolcic+2015 & Vd-17871 & 4.622 & 9.5$\pm$2.0 & 12.3$\pm$2.4 & 13.0$\pm$3.7 & 13.01 & 1.0   \\
  Sklias+2014\tablefootmark{b} & A68/nn4 & 3.19 & 45.3$\pm$1.87 & 38.04$\pm$2.19 & 17.74$\pm$1.5 & 13.01 & 2.3   \\
  Bothwell+2013 & SPT0538-50 & 2.8 & 326.0$\pm$23.0 & 396.0$\pm$38.0 &
  325.0$\pm$24.0 & 14.09 & 21.0 \\ 
  Magdis+2011 & GN20 & 4.055 & 18.6$\pm$2.7 & 41.3$\pm$5.2 & 39.7$\pm$6.1 & 13.34 & 1.0   \\
  Cox+2011 & HATLASID141 & 4.24 & 115.0$\pm$19.0 & 192.0$\pm$30.0 & 204.0$\pm$32.0 & 14.1 & 20.0 \\
  Conley+2011\tablefootmark{b} & HLSW-01 & 2.957 & 425.0$\pm$10.0 & 340.0$\pm$10.0 & 233.0$\pm$11.0 & 14.22 & 11.0 \\
  Combes+2012 & HLSA773 & 5.24 & 85.0$\pm$8.0 & 168.0$\pm$8.0 & 203.0$\pm$9.0 & 14.26 & 11.0 \\
  Reichers+2013\tablefootmark{b} & HFLS3 & 6.34 & 12.0$\pm$2.3 & 32.4$\pm$2.3 & 47.3$\pm$2.8 & 13.79 & $<$3.5   \\

\hline
\end{tabular}
\tablefoot{
\tablefoottext{a}{\rm IR luminosity derived using DH02 template fitting,
  uncorrected for lensing magnification.}
\tablefoottext{b}{Calibration errors and confusion noise are not
  included.}
}
\label{tab:smglist}
\end{table*}

 We select from the literature a sample of 57 galaxies at
 $z\,>\,2.5$. We attempt to include all
  galaxies that are spectroscopically confirmed and have
  detections in all SPIRE bands in this sample, excluding those with
  apparent signatures of active galactic nucleus (AGN). The galaxies and
  their sources are listed in Table \ref{tab:smglist}. Since a
  considerable number of these galaxies are lensed, the magnification
  factor of each galaxy is also listed in Table \ref{tab:smglist}.
  The galaxies in this sample are taken from different works, and
  therefore are inhomogeneous in selection methods as well as redshift
  measurements. We shortly describe each work in the following.

\textsl{\citet{weiss2013}}: The SPT sample from \citet{weiss2013} includes $1.4$mm selected
galaxies in the South Pole Telescope Sunyaev-Zeldovich (SPT-SZ) survey
\citep{carlstrom2011}. Synchrotron dominated sources classified by \citet{vieira2010} 
have been removed from this sample. The spectroscopic redshifts of these galaxies
are derived from the Atacama Large Millimeter/submillimeter Array
(ALMA) observations. Sixteen of these sources are at $z>2.5$ and
  have SPIRE detections in all three bands. For three out of the 16
galaxies, \citet{hezaveh2013} derived the IR luminosities using the
MBB model with $\beta=2.0$, and their lens magnification factors from
lensing models. There is no measurement of the $L_{\rm IR}$ or
  lens magnification factor available in the literature for the left
sources. We assume no lens magnification for these galaxies in Table
\ref{tab:smglist}, however, these galaxies are likely to be lensed.

\textsl{\citet{magnelli2012}}: \citet{magnelli2012} collected a
  sample of submillimeter galaxies with
  spectroscopic redshift and Herschel observations. Sources 
  with no PAH features are considered to be AGN and were excluded 
  from this sample. We select 11
  galaxies in their sample with $z>2.5$ and detections in three SPIRE
  bands. In their work, the $L_{\rm IR}$ is calculated from
  multitemperature SED fitting. 

 \textsl{\citet{casey2012}}: Six out of the 36 \textit{Herschel} sources
  of the sample
\citet{casey2012} are at $z\ >\ 2.5$ and have no sign of AGNs according to their optical spectra.
The redshift of these galaxies was spectroscopically confirmed with
the Keck I Low Resolution Imaging Spectrometer (LRIS) and the Keck II DEep
Imaging Multi-Object Spectrograph (DEIMOS). The $L_{\rm IR}$ are
obtained fitting a single MBB with $\beta=1.5$ and a
mid-infrared (MIR) power law with a slope of $\alpha=2.0$ to the IR data.

\textsl{\citet{saintonge2013}}: \citet{saintonge2013} present a sample
of lensed galaxies at $z=1.4$
{--} $3.1$ with \textit{Herschel}/SPIRE data. These galaxies have
intrinsic properties similar to ultraviolet (UV)-selected 
lyman break galaxies (LBGs). A comparison of
these lensed galaxies with unlensed galaxies suggests that they are
representative of star-forming galaxies at their
redshifts. The [NII]/H$\alpha$ ratios of these galaxies are in the range 
expected from star-forming galaxies, indicating that they do not harbor AGNs. Six of
these galaxies are at $z > 2.5$ and have detection in each
SPIRE band. We remove one galaxy (J1527) from our sample because its
SPIRE photometry has very large uncertanties
($\sigma(S_{\nu})/S_{\nu} \sim 10$) and is not consistent with its
PACS measurement (see their Figure 5). The IR luminosities of these
galaxies given by \citet{saintonge2013} are calculated from the SED
fitting using \cite{dl2007} models.

\textsl{\citet{dowell2014}}: \citet{dowell2014} selected a sample of
`$500$ $\mu$m risers', which
are high-z galaxies, as confirmed by the follow-up observations with
CARMA. The IR luminosities for these galaxies are calculated by
fitting MBB models (leaving $\beta$ as a free parameter) to the SPIRE
photometry and assuming no lensing magnification. We assume the AGN 
contribution is  negligible.

\textsl{\citet{roseboom2012}}: Four of the submillimeter sources of 
\citet{roseboom2012} have detections in all three SPIRE bands. 
They used an MBB model with $\beta=1.8$ and
constant optical depth to fit the SED of the sources and derive their
IR luminosity, assuming AGN contribution is negligible. 

{\sl \citet{negrello2010}}: The two galaxies of \citet{negrello2010},
SDP81 and SDP130, are selected using a flux density cut at 500\,$\mu$m
($S_{500}\ > 100$ mJy). They are strongly lensed with a magnification
factor of 19 and 6, respectively. Their total intrinsic  $L_{\rm IR}$
are derived by \citet{negrello2010} from multiwavelength SED fitting
using models from \cite{des2008}. The MIR emission of the two 
galaxies suggests little AGN contribution \citep{negrello2014}.

\begin{figure}
\centering
\includegraphics[width=\linewidth]{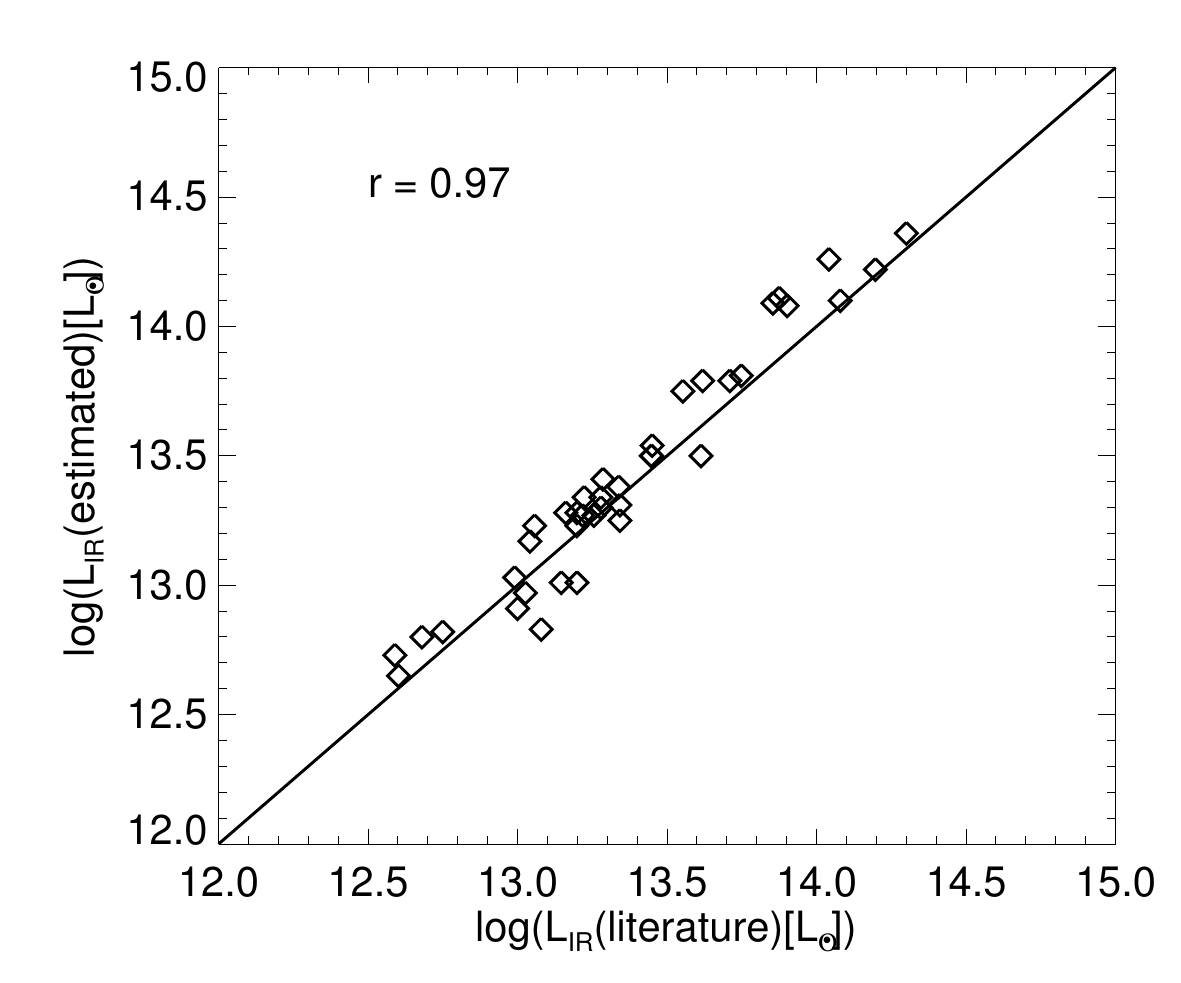}
\caption{Comparison of the observed IR luminosity $L_{\rm IR}$
  derived from SED fitting based on the templates of DH02
  and the values from literature.  }
\label{fig:est_ir}
\end{figure}

\textsl{\citet{smolcic2015}}: AzTEC1 and Vd-17871 have
  $z>2.5$. The redshift of AzTEC1 is confirmed to be $4.34$ by SMA
  observation (Yun et al. {\sl in prep.}). The redshift of Vd-17871 is
  obtained by DEIMOS/VLT follow-up observation. The $L_{\rm IR}$ given
  in this work is calculated by fitting the SEDs using \citet{dl2007}
  models. The widths of $Ly\alpha$ lines of the two galaxies 
  rule out the possibility that they are powerful AGNs.

\textsl{\citet{sklias2014}}: We select A68/nn4 from the work of
  \citet{sklias2014}, who present Herschel observations for several
  lensed star-forming galaxies with well-determined spectroscopic
  redshifts at $z=1.5-3$. The object A68/nn4 is  at $z=3.19$. The $L_{\rm
    IR}$ in this work is calculated from SED fitting using \citeauthor{ce2001} (2001;CE01) templates.

\textsl{SPT0538-50}: SPT0538-50 is a star-forming galaxy first
  discovered by the South Pole
Telescope. The redshift is measured by \citet{aravena2013} from CO
observation. \citet{bothwell2013} present the Herschel observation for
this galaxy and calculated the IR luminosity using
CIGALE\footnote{Code Investigating GALaxy Emission,
  http://cigale.lam.fr, \citep{noll2009}} SED fitting. The \citeauthor{dh2002} (2002;DH02) models are used for IR part.

\textsl{GN20:} GN20 is first detected by \citet{pope2006} with a redshift
$z=4.05$ obtained by \citet{daddi2009}. The SPIRE fluxes were obtained by
\citet{magdis2011}, who also estimated the IR luminosity using the
models of \citet{dl2007}.

\textsl{H-ATLAS ID141}: H-ATLAS ID141 is a source lying at $z=4.24$ with an apparent IR
luminosity of $\sim$ $(8.5\pm0.3)\times 10^{13} \mu^{-1}
L{\odot}$ obtained from an MBB fitting, under the optically
thin assumption, and a magnification factor between 10 and 30
\citep{cox2011}. 

\textsl{HLSW-01}: HLSW-01 (HERMES J105751.1+573027) was
spectroscopically confirmed at $z=2.96$ by \citet{riechers2011}. This
multiple source is lensed by a group of galaxies resulting in a
magnification factor of $\mu=10.9\pm0.7$ \citep{gavazzi2011}.
The SPIRE fluxes of HLSW-01 are from the HerMES SCATv3.1
catalog \citep{smithaj2012}, and its IR luminosity is calculated by
\citet{conley2011} using SED fitting with an MBB model.

\textsl{HLSA773}: HLSA773 is a source at $z=5.24$ with an IR luminosity of
$1.1\times10^{14}/\mu$ $L_{\odot}$ obtained from an optically thin MBB
model with $\beta=2$, and $\mu$ estimated at $\sim$ $11$
\citep{combes2012}. 

\textsl{HFLS3}: HFLS3 satisfies $S_{500}>S_{350}>S_{250}$, has a
redshift of $z=6.34$, and a $L_{\rm IR}$ of $4.16\times10^{13}$
$L_{\odot}$ estimated from an MBB SED fitting
\citep{riechers2013}. According to \citet{cooray2014}, the lensing
  magnification of this galaxy is less than 3.5.

The IR luminosities from the literature are derived using different
  wavelength ranges and different SEDs. To present more
  uniform measurements of IR luminosities, we recalculate $L_{\rm IR}$
  by fitting the SEDs using DH02 templates and the three-band SPIRE
  fluxes. The results are shown in Table \ref{tab:smglist} (the values
  are not corrected for lensing magnification). We compared the
  results with the literature values in Figure
  \ref{fig:est_ir}, and found that our results are in agreement with
  the estimations given in the literature.

These high-z galaxies are bright in IR, suggesting that they
are starbursting galaxies. However, at high-z the
definition of main-sequence (MS) galaxies and starbursts (SBs) is not clear.
Several works have shown that the extremely high star formation rate (SFR) of high-z
sources is due to the large amount of cold gas, and that SBs
should be defined according to the compactness of the
star-forming region instead of the SFR \citep{elbaz2011}. In the case of our sample
galaxies, further information is needed to be able to
state if they belong to the MS or not. Therefore, we do not
distinguish MS sources from SBs when we compare them
with SED libraries in Section \ref{subsec:data_sed}.

In our sample we have removed galaxies with apparent AGN features, however, we cannot rule out galaxies with less powerful or obscured AGNs. Moreover, 
the AGN identification is not uniform due to the inhomogenous data used 
in each work. As a result, some galaxies in our sample may have embedded AGNs.
In Section \ref{subsec:agn} we  show that the AGN contribution is 
insignificant for the wavelength range discussed in this work.

\subsection{Comparison samples}
\label{subsec:compsamp}

In addition to the high-z sample of identified objects,
we use other samples for comparison.
First, a UV-selected sample at $z\sim3$ and $4$ compiled
by \citet{heinis2014} from the Herschel Multi-tiered Extragalactic
Survey key program (HerMES,
\citealt{oliver2012}). The redshifts of these galaxies are
  taken from the photometric redshift catalogue built from COSMOS data
  by \citet[][version 2.0]{ilbert2009}, and the mean error of the
  redshift is about $0.4$ and $0.7$ at $z\sim3$ and $z\sim4$, respectively. The IR fluxes are
estimated from HerMES data using stacking analysis. The IR luminosity $L_{\rm IR}$ of the stacked sample is
$\sim$ $4\times10^{11}L_{\odot}$. We use the UV-selected galaxies  to
compare with the spectroscopically confirmed high-z sample in Section
\ref{subsec:uvsel}.

When discussing the methods to select high-z populations using
the SPIRE color-color diagram (Section \ref{sec:photoz}), we use a sample of $500\,\mu$m
selected galaxies taken from \citet{rr2014}. These sources are
obtained by matching the HerMES SCAT catalog
\citep{oliver2012,wang2014} in the SWIRE-Lockman area
\citep{lonsdale2003} with the SWIRE catalog \citep{rr2008}, with the
requirement that the $24\mu$m fluxes are brighter than $100\mu$Jy.  
The redshifts are derived using up to 13 photometric bands available
in the Lockman area. When there are multiple counterparts in the SWIRE
catalog, the source with the highest $450\mu$m predicted according to
the SED is selected. The sample include 1335 SPIRE $500\mu$m selected
sources, 967 of which have SWIRE associations and 783 sources are with
detections in all the three SPIRE bands. The redshifts of the sample
range from $z\sim 0$ to $z\sim 5$. The photometric
redshift error for this sample is about $4\%$ in $(1+z)$ 
\citep{rr2014}. The selection effect and the reliability of the redshifts
for this sample are discussed in Section \ref{sec:photoz}.

\section{SED tracks on the SPIRE color-color diagram}
\label{sec:sedtemp}

\subsection{Single-temperature modified blackbody models}
\label{subsec:mbb}

\begin{figure}
\centering
\includegraphics[width=\linewidth]{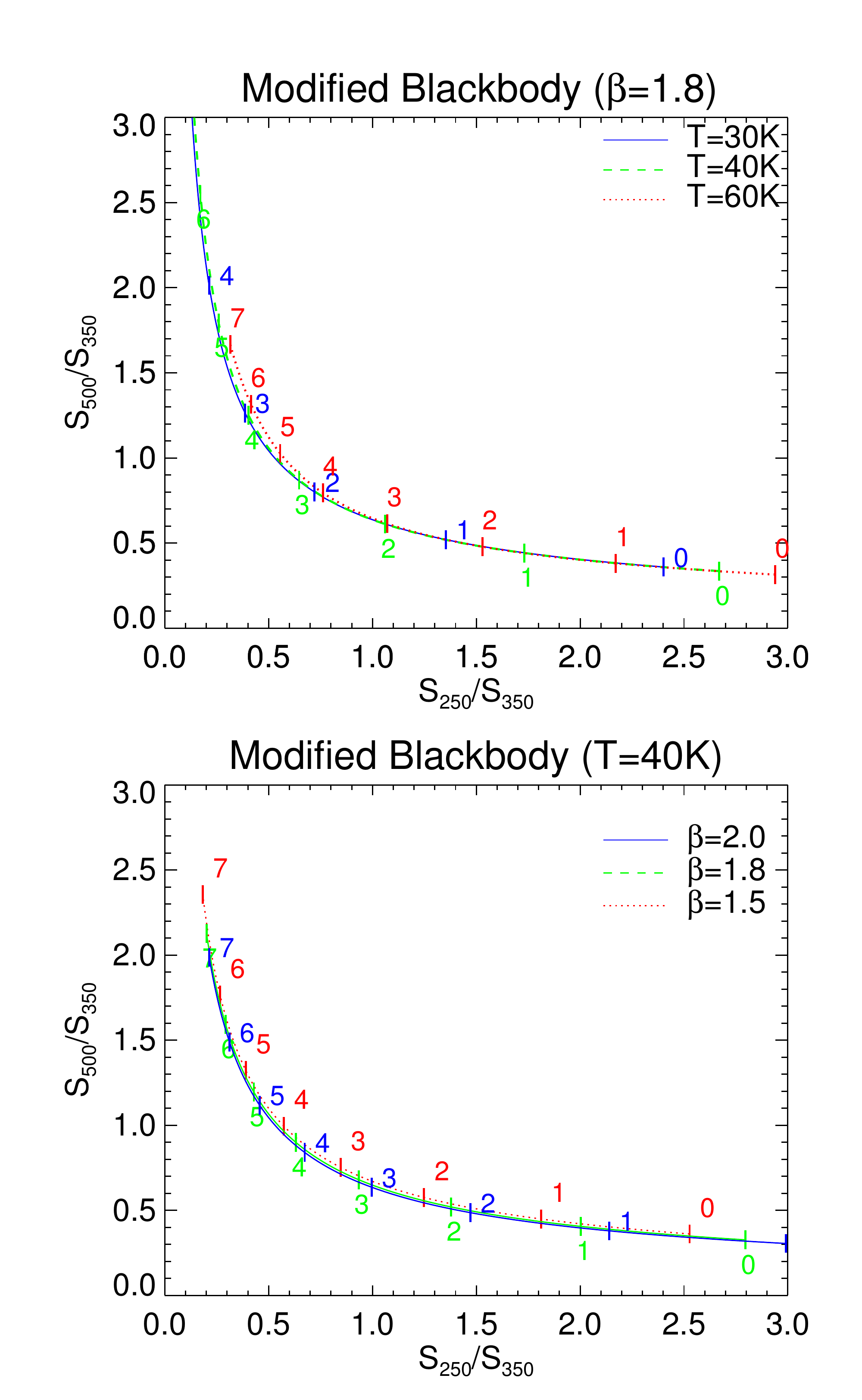}

\caption{The SPIRE color-color relation for MBB models. Upper
  panel: MBB models with $\beta=1.8$, $\nu_0=c/100 \mu$m and $T_{\rm
    d}$=30\,K (Blue solid line), 40\,K (green dashed line), and 60\,K
  (red dotted line). Lower panel: MBB models with $T=$40\,K and
  $\beta=2.0$ (Blue solid line), 1.8 (green dashed line), and 1.5
  (red dotted line). The numbers on the plot indicate the redshifts.}
\label{fig:mbb_ill}
\end{figure}

We use an MBB model to derive the dust temperature of the galaxies
of our sample, as commonly done in the literature
\citep[e.g.,][]{roseboom2012,casey2012,huang2014}, with
\begin{equation}
f_{\nu}=(1-e^{-\tau_{\lambda}}) B_{\nu}(T_d)
,\end{equation}
\noindent where the optical depth $\tau_{\lambda} \propto (\nu/\nu_0)^{\beta}$, ${\beta}$ is the grain
emissivity index with values ranging from $\sim$ 1.5 to $\sim$ 2
\citep{hildebrand1983}, $\nu_0$ is the frequency at which the radiation becomes
optically thick, and $B_{\nu}(T_{d})$ the blackbody
spectrum at a dust temperature of $T_{\rm d}$. The upper panel of Figure
\ref{fig:mbb_ill} shows the SPIRE color-color diagram of the MBB
model with dust temperatures at 30, 40, and 60\,K. Here, $\nu_0$
is assumed to be $c/100\,\mu$m and $\beta=$1.8.
We follow the results of \citet{roseboom2012},
who found that a $T=40$K MBB model with $\nu_0=c/100 \mu$m
and $\beta=1.8$ can adequately describe the color
of their sample of galaxies with $2<z<5$. The upper panel of Figure \ref{fig:mbb_ill}
shows that, at the same redshift, cold dust galaxies are
above and more to the left than hot dust galaxies, i.e., at higher $S_{500}/S_{350}$
and lower $S_{250}/S_{350}$ compared to hot dust galaxies.
Therefore, we refer to the upper-left part of the SPIRE color-color diagram
as the `cold region', while the lower-right part of the diagram is the `hot
region' according to the MBB models. The lower panel of Figure
\ref{fig:mbb_ill} shows that taking different value of
$\beta$ can cause systematic differences when estimating the dust
temperatures. However, this does not affect our conclusions since we
are only interested in the relative dust temperatures. In the
  following part, we assume $\nu_0=c/100\,\mu$m and $\beta=$1.8 for
  MBB models. 

\subsection{SED libraries}

\begin{figure*}
  \centering
  \includegraphics[width=0.45\linewidth]{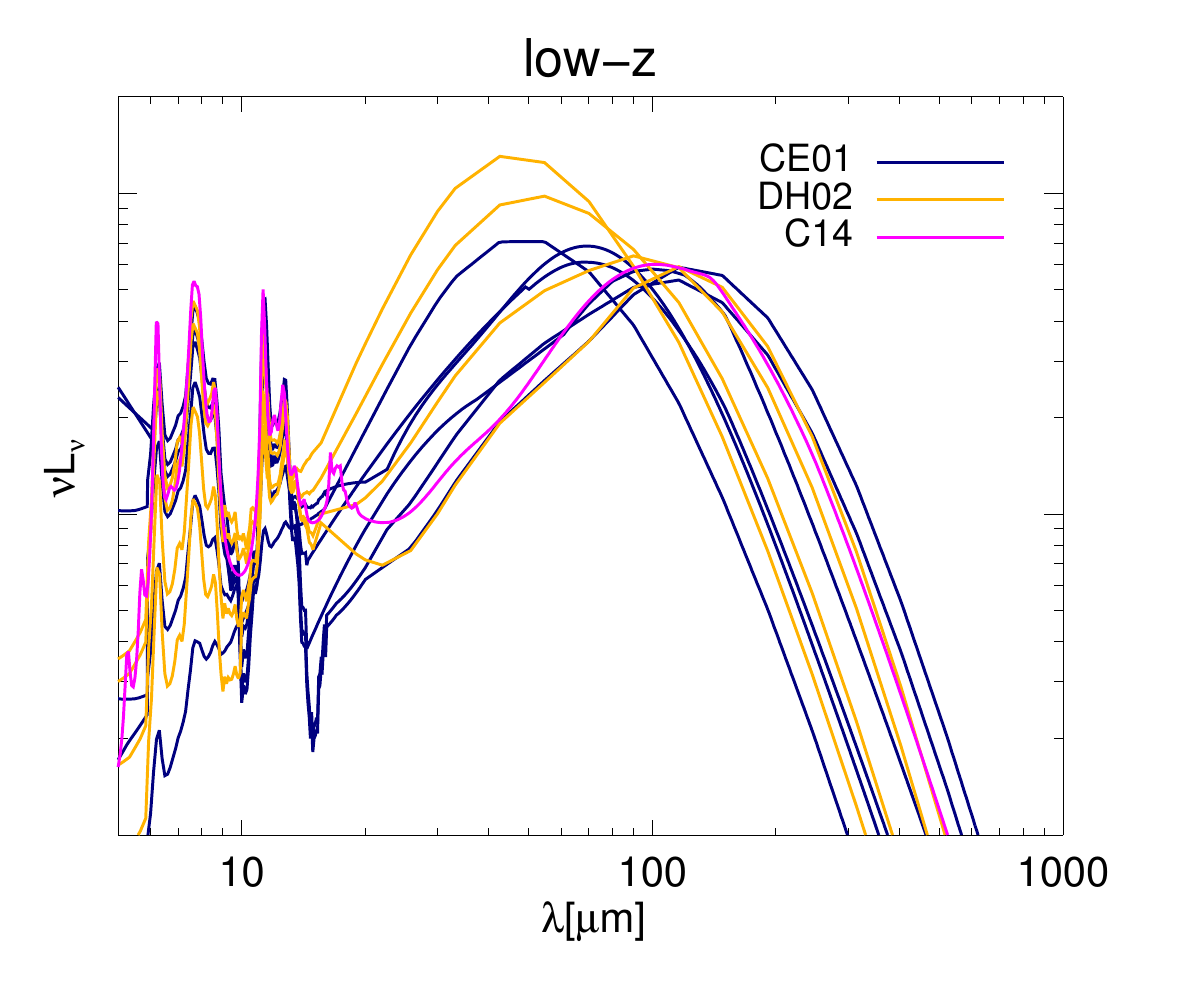}
  \includegraphics[width=0.45\linewidth]{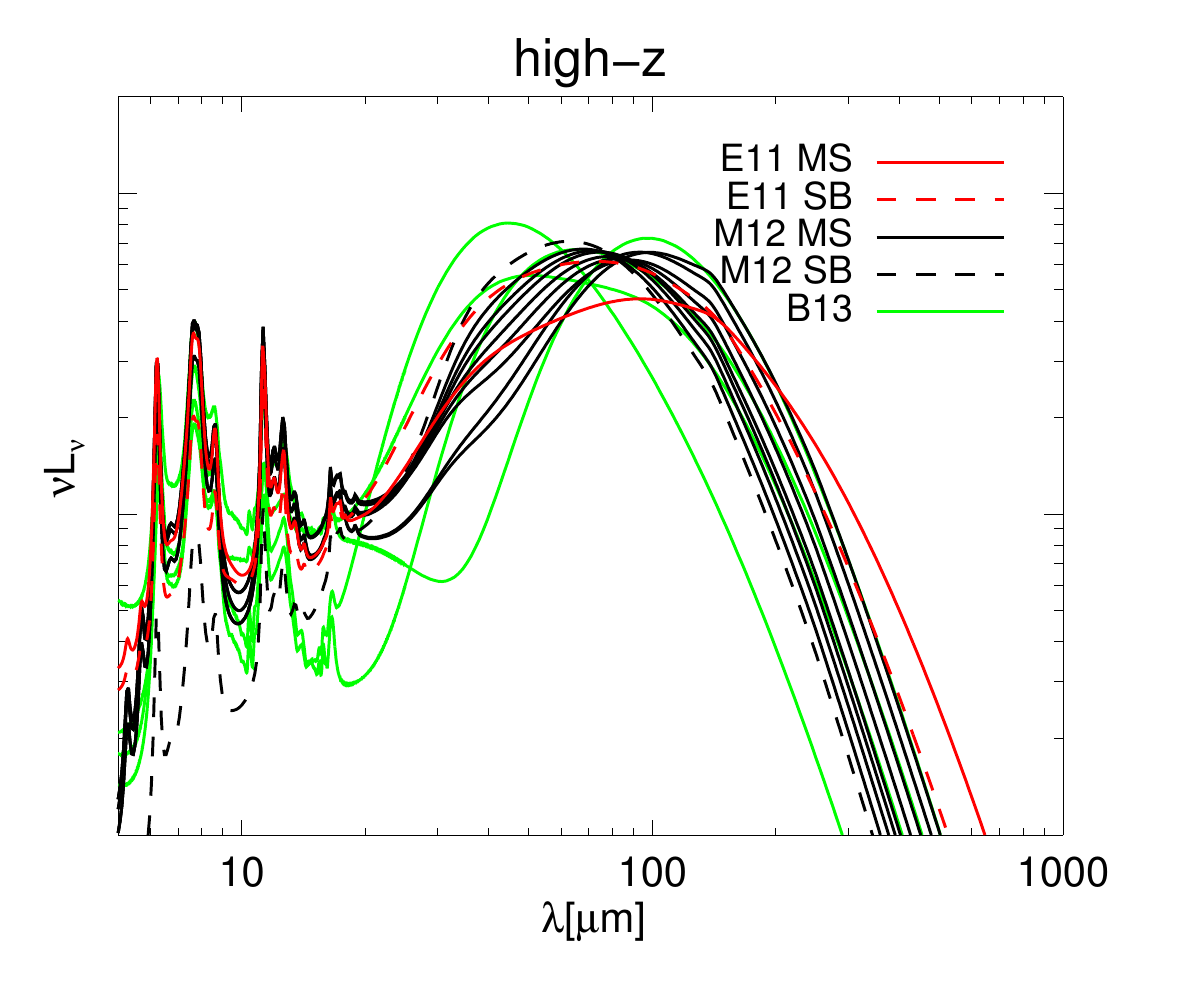}
  \caption{SED templates studied here. Left panel: low-z
    templates. In navy:  CE01 templates with $L_{\rm IR}$ $\sim$
    $10^{9}$, $10^{10}$, $10^{11}$, $10^{12}$, $10^{13}$
    $L_{\odot}$. In yellow:  DH02 templates with $\alpha$ $=$
    $1.0$, $1.5$, $2.0$, $2.5$. In magenta:  C14 mean
    template. Right panel: high-z templates. In red:  E11
    templates. The solid line denotes the MS template and the dashed
    line denote the SB template. In black:  M12 templates. The
    solid line denotes the MS template and the dashed line denotes the
    SB template. In green: four SF templates selected from B13.}
  \label{fig:seds}
\end{figure*}

The SED libraries used in this work are CE01,
DH02, \citeauthor{elbaz2011} (2011; E11), \citeauthor{magdis2012}
(2012; M12), \citeauthor{berta2013} (2013; B13), and 
\citeauthor{ciesla2014} (2014; C14), shown in
Figure \ref{fig:seds}. Most of these libraries are based on data at
$z<3$. Currently, no library exists for $z>3$ objects. Therefore, we
need to extrapolate these templates to higher redshift  to
compare with the data. According to M12, at $z>3$ the SED of a MS galaxy stops evolving because of the flattening of the evolution
of the specific star formation rate (sSFR). Therefore, we assume no
evolution after $z>3$ for these libraries. For $z<3$, only the M12
MS library takes  the evolution
of IR SEDs with redshifts into account, so we set the templates of M12 as the
reference and compare all other templates to them. The color-color
diagrams for the templates selected from these libraries are shown in
Figure \ref{fig:ccsed}. In the following, we introduce the libraries
and discuss the position of the templates on the SPIRE color-color
diagram.

\subsubsection{Pre-\textit{Herschel} templates}

Pre-\textit{Herschel} libraries, such as CE01 and DH02, are
commonly used in the studies of high-z galaxies
\citep[e.g.,][]{magdis2010,elbaz2011,berta2013,riechers2013,heinis2014}. The
templates of CE01 are built to reproduce the ISO, IRAS, and SCUBA
observations, using the dust models of \citet{silva1998} and ISO
MIR spectra. These templates are parameterized by the total IR
luminosity, ranging from $L_{\rm IR}\sim 10^{8}$ to $\sim10^{13}$
$L_{\odot}$. The templates of DH02 are based on IRAS, ISO, SIRTF, and
SCUBA observations. The local SEDs based on the models of \citet{dbp1990}
are combined assuming the distribution of dust mass $M_{\rm d}$ as $dM_{\rm d}\
\propto\ U^{-\alpha}\,dU$, where $U$ is the intensity of the radiation
field and the exponent $\alpha$ ranges from 1 to 2.5 for star-forming
galaxies. Galaxies with small $\alpha$ values are more actively
forming stars.

\subsubsection{Post-\textit{Herschel} templates}

Based on GOODS-\textit{Herschel} data at $0\ <\ z\ <\ 2.5$ and local
data of AKARI, ISO, and IRAS, \citet{elbaz2011} found that most ULIRGs
at high-z are actually forming stars in a ``quiescent'' mode similar
to local MS galaxies, and their high SFR are due
to the large amount of molecular gas. Therefore, they suggest
universal SED templates for MS galaxies and for SB galaxies without
assuming any evolution. The templates were built by fitting their
sample ($0<z<2.5$), using a ``diffuse ISM'' component and a
``star-forming region'' component. The models of the two components
are given by \citet{galametz2009} and \citet{galliano2011}. The galactic dust properties are adopted from
\citet{zubko2004}. The distribution of dust mass heated by different
starlight intensities $U$ is assumed to be a power law: $dM_{\rm
  d}/dU\ \propto\ U^{-\alpha}$.

M12 derive the MS SEDs by fitting the \citet{dl2007} model to the data
of a well-defined set of normal MS galaxies, including nine normal
star-forming galaxies at $z$ $\sim$ $0.5$ and $\sim$ $1.5$, three
high-z SMGs, and a stacked sample of $>$ $4000$ MS galaxies at $0.5\
<\ z\ <\ 2.5$, all of which have spectroscopic redshifts, rich
rest-frame UV to MIR photometry and \textit{Herschel} detection. They find
that SEDs of MS galaxies can be characterized by the dust-weighted mean
starlight intensity scale factor $<U>$, which is proportional to
$L_{\rm IR}/M_{\rm d}$, and is mildly evolving with cosmic time as
$<U>\propto\,(1+z)^{1.15}$. They build eight SED templates (U1 to U8)
for MS galaxies at eight redshift bins (Table \ref{tab:magdis}). In
addition, they suggest that SB galaxies have a universal SED,
and use the fit of GN20 ($z\ \sim\ 4$) as the template for SB galaxies.

\begin{table}[h!]
\caption{Redshifts range for M12 models}
\begin{center}
\begin{tabular}{cc}
MS templates & z \\
\hline
U1& 0 -- 0.025\\
U2& 0.05 -- 0.275\\
U3& 0.3 -- 0.625\\
U4& 0.65 -- 0.975\\
U5& 1.0 -- 1.30\\
U6& 1.325 -- 1.725\\
U7& 1.75 -- 2.25\\
U8& 2.27 --3.0\\
\hline
\end{tabular}
\end{center}
\label{tab:magdis}
\end{table}

\citet{berta2013} built 32 SED templates from PEP and HerMES
data. Each template is the median SED of a class of galaxies
defined from the multiwavelength color space. Eight of these SED templates
are for rare objects (outliers of the main classes defined from the color
space), including torus, AGNs or even elliptical galaxies. The SEDs
are fitted using MAGPHYS code \citep{des2008}.

Other post-\textit{Herschel} templates are derived from local observations
\citep[e.g.,][C14]{smith2012}. We use the template
of C14 to compare with the high-z
libraries. Using \citet{dl2007} models, C14 fit the UV to submillimeter SEDs of a
sample of 322 nearby galaxies, the \textit{Herschel} Reference Survey
\citep[HRS,][]{boselli2010}, which is volume limited and selected in K-band.
The fitted SEDs are averaged to provide the mean SED template.

\subsection{SPIRE color-color diagram based on different SED
  libraries}

\begin{figure*}
\centering
\includegraphics[width=0.85\linewidth]{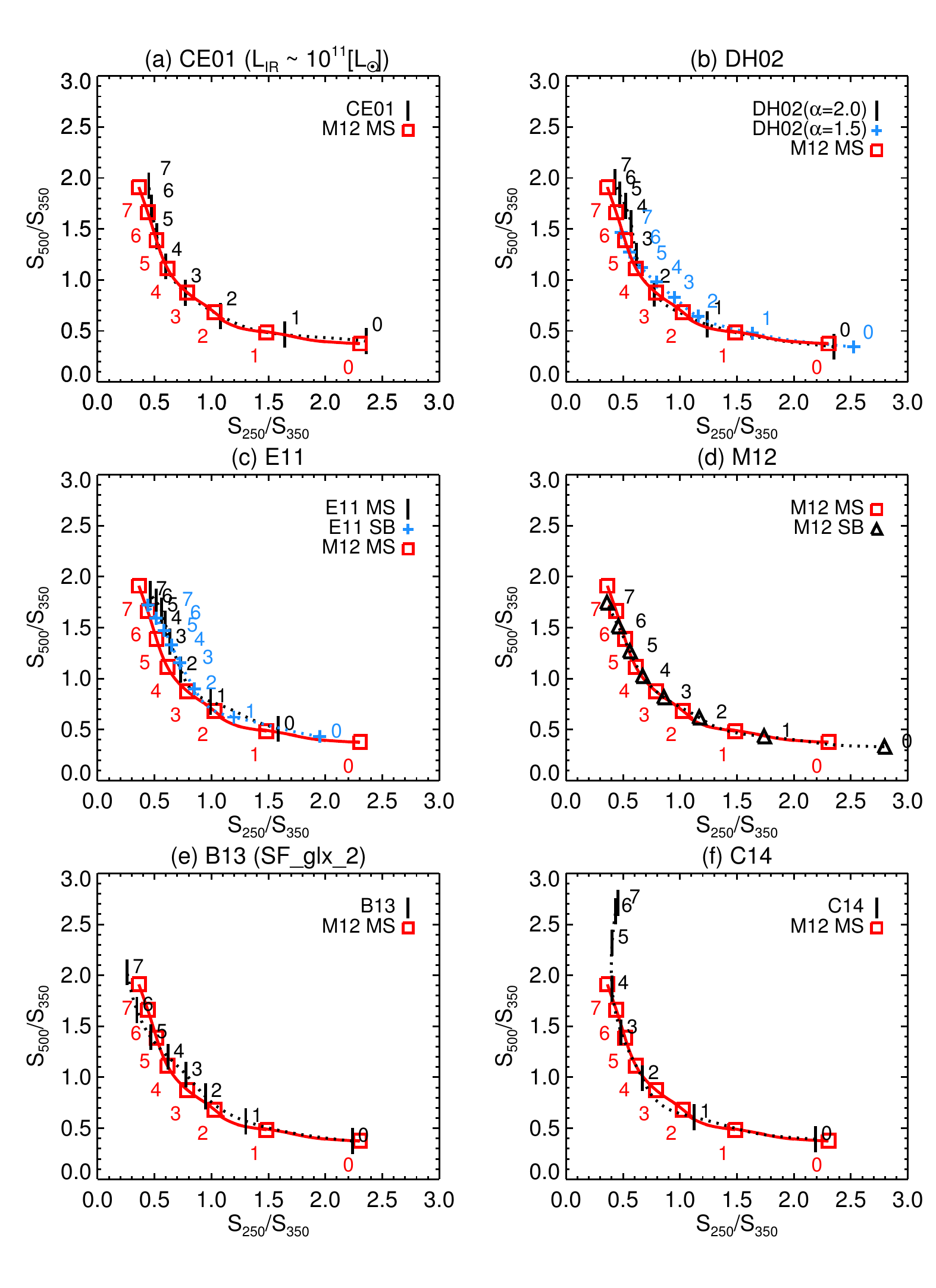}
\caption{SPIRE color ratios based on different SED libraries. The
  numbers on the plot indicate the redshifts. The SPIRE color ratios
  based on the M12 templates are plotted in each
  panel as a reference (red line with squares).  Panels (a): 
  template of CE01 with $L_{\rm IR}\sim10^{11}L_{\odot}$;
  (b):  template of DH02 with $\alpha=2.0$ (black) and
  $\alpha=1.5$ (blue); (c) 
  templates of E11 for MS galaxies (black)
  and SBs (blue); (d)  templates of M12 for
  MS galaxies (red) and SBs (black); (e) 
  SF\_glx\_2 template of B13, which is derived from
  star-forming galaxies in COSMOS field; (f) template of
  C14. }
\label{fig:ccsed}
\end{figure*}

Figure \ref{fig:ccsed} shows the SPIRE color-color diagram predicted
by different SED libraries (see more color-color diagrams in the
Appendix). The M12 MS templates are plotted in each panel for
comparison. At $z < 3$, the U1 to U8 templates are used for the
redshift bins, which they are assigned to by M12. Beyond $z=3$, the U8
template is used to derive the colors considering the flattening of
the evolution of the sSFR.

Panels (a) and (b) of Figure \ref{fig:ccsed} present the SPIRE
color-color diagram based on the pre-\textit{Herschel} templates. Panel (a)
shows that the template of CE01 with $L_{\rm IR}$ $=$
$10^{11}L_{\odot}$ produces a SPIRE color-color relation
similar to the MS templates of M12. Panel (b) shows that in the DH02
library, the template with $\alpha\ =\ 2.0$ agrees well with the MS
templates of M12 in the local universe, whereas as $z$ increases to
$2$, $\alpha$ must reduce to $\sim$ $1.5$ to be consistent with the colors
reproduced by the M12 MS templates. Beyond $z=2$, a template of DH02
with $\alpha$ value between $1.5$ and $2.0$ seems to be in agreement
with the MS template of M12. Panels (a) and (b) of Figure
\ref{fig:ccsed} also show that the templates of CE01 and DH02 do not
produce $S_{500}/S_{350}$ values as large as predicted by the MBB models.

Panel (c) shows the comparison between the templates of E11 with the
MS templates of M12 on color-color diagram. Both the MS
and SB templates of E11 predict colder colors than M12 galaxies at $z
< 6$. The difference is due to the difference of the Rayleigh-Jeans
side between E11 and M12 SEDs (Figure \ref{fig:seds}).

Panel (d) compares the M12 MS template to their SB template, i.e., the
SED of GN\,20. It appears that at every redshift the SB template is
located in hotter regions of the color-color diagram compared to the
MS templates. This is not surprising since the star-forming activity
in SB galaxies is more intense than in MS galaxies and can thus heat
the dust more efficiently.

\clearpage

We select four star-forming galaxy templates from B13.
We found that, on a color-color diagram, the
template SF\_glx\_2 (derived from star-forming galaxies of the COSMOS field)
reproduces SPIRE colors in agreement with M12
MS templates at $z>4$ (Panel [e] of Figure \ref{fig:ccsed}). The
SF\_glx\_1 template (star-forming galaxy SED derived from the GOODS-S field)
reproduces much colder colors than the M12 templates. The MIRex\_SF\_glx
(MIR excess star-forming galaxies) and Obs\_SF\_glx (obscured
star-forming galaxies) templates have hotter SPIRE colors than the M12
MS templates (see Figure \ref{fig:berta_sel}).

Panel (f) shows that at $z\ \sim\ 0$ the C14 template agrees well with
the M12 MS template (U1) on the SPIRE color-color diagram, whereas at
high-z the SPIRE colors of the C14 template is much colder than the M12
template.

Many studies have shown that at $\lambda > 100 \mu$m the
  SED can be well approximated by a simple MBB
  \citep[e.g.,][]{gordon2010,davies2012}. Indeed, most of the
  templates can be well parameterized by a single temperature MBB at the SPIRE wavelength. Therefore, MBB models can
  be used for simplicity. However, for models with redshift evolution
  (e.g., M12) ,or those with wide peaks (e.g., E11), MBB is
  not a good approximation. Also, when going
  to higher redshifts ($z\gtrsim 2$), the observed SPIRE bands are moving to
  shorter wavelength where the dust emission becomes more complicated
  so that the MBB model is over simplified.

\subsection{Contribution from AGNs }
\label{subsec:agn}

Some libraries include contributions from AGNs, however, several works showed that the AGN contribution is negligible
\citep{elbaz2010,elbaz2011,casey2014} at long wavelengths
(longer than $\sim$30\,$\mu$m). Furthermore,
\citet{hatziminaoglou2010} showed that the SPIRE colors of AGNs and
non-AGN, star-forming galaxies are indistinguishable. We also find
that the host galaxy dominated AGN template of B13
generates SPIRE colors very similar to M12 templates. To
include an AGN contribution, \citet{dale2014} provided a modified version
of DH02 templates in which the contribution of a quasar of different
intensity can be added. Comparing the \citet{dale2014} template containing
an AGN emission with its corresponding DH02 template using
$\alpha=2.0$, we find that when the 5-20 $\mu$m AGN fraction is less
than $50$\%, the difference of SPIRE colors $(S_{250}/S_{350},\
S_{350}/S_{500})$ given by the two templates are less than
$(0.1, 0.15)$. In this paper, since we are working with SPIRE colors
where the AGNs hardly affect the results, we do not discuss the
possible AGN contamination.

\subsection{CMB contribution}

At high redshift the impact of cosmic microwave background (CMB) on the
observed IR SED can be significant. The CMB affects the observed
dust emission with two competing processes: it heats the ISM in galaxies
and  produces a considerable background against which galaxies are
observed \citep{cunha2013}. The CMB effect can complicate the comparison
between data and templates. Fortunately, according to
the results of \citet{cunha2013}, for the redshift range considered here
($z<6.5$) the effect of CMB on dust temperatures is $\sim \pm 3$K. For
galaxies with lower intrinsic dust temperatures (e.g., 18K), the CMB
effect on dust temperatures is a function of redshifts. But for
galaxies with higher intrinsic dust temperatures (e.g., 40K), the
effect is almost constant at different redshifts. In Section
\ref{sec:res} we  find that our high-z galaxies have dust
temperatures higher than 30K. Therefore, the CMB effect on these
galaxies is negligible for our discussion in this paper.

\section{Results}
\label{sec:res}
\subsection{Galaxies at $z>2.5$ on the SPIRE color-color diagram}
\begin{figure*}[!htb]
  \centering
  \includegraphics[width=\linewidth]{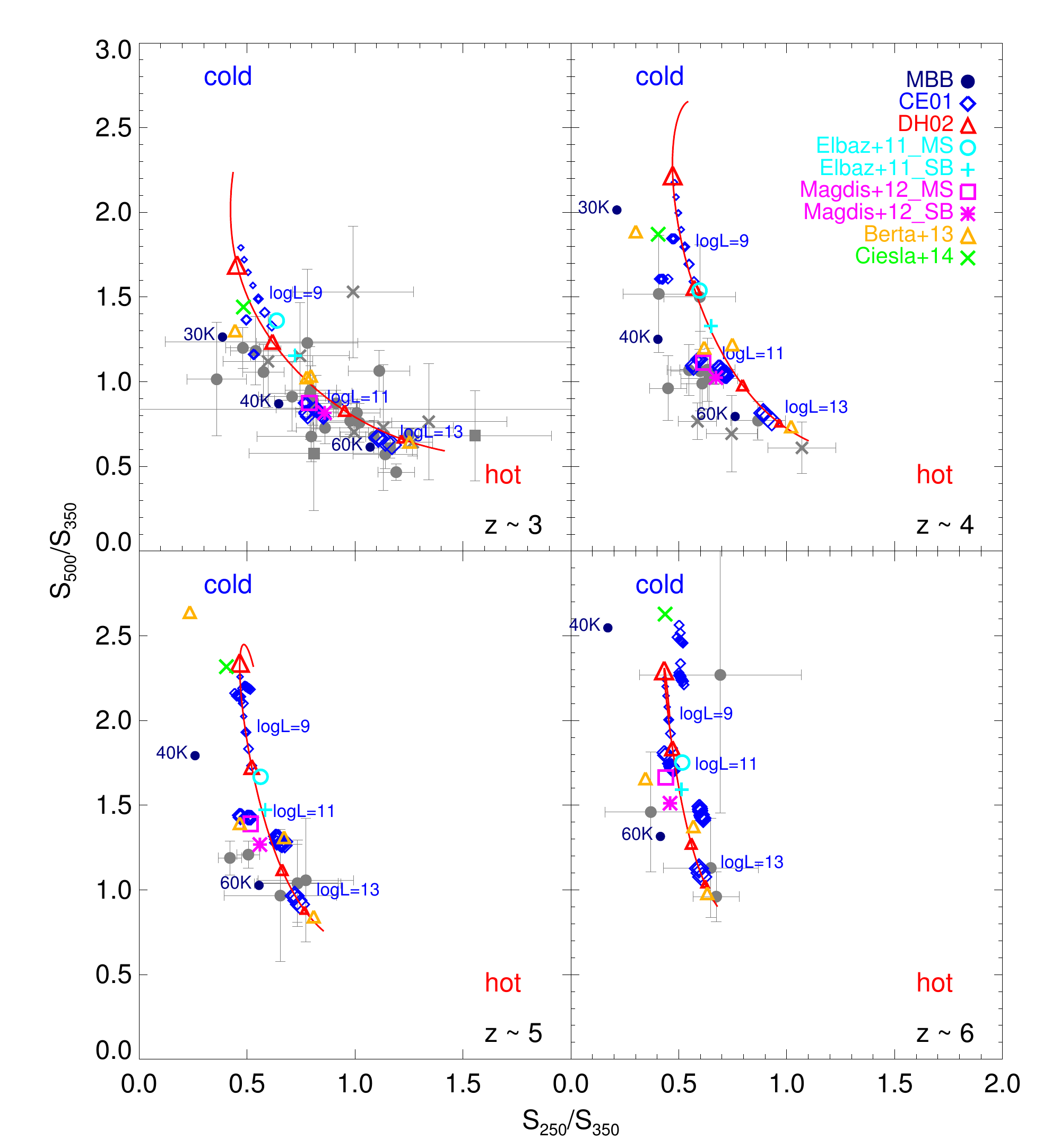}  
  \caption{Comparison of the high-z observations and the SED templates
    on the SPIRE color-color diagram in different redshift bins ($z$
    $\sim$ $3$, $4$, $5$ and $6$). The observations data are in gray
    and the results of the templates are in color. The SED templates
    plotted in the figure include: 1) MBB models with $T=30$, $45$ and
    $60$K (navy dots); 2) templates of CE01 (blue open diamonds,
    the sizes of which are increased with the IR luminosity);
    3) templates of DH02 (red line) with $\alpha$ $=$ $1.0$,
    $1.5$, $2.0,$ and $2.5$ marked in triangles with increasing sizes;
    4) MS and SB templates of E11 (cyan circles and
    plus signs); 5)  MS and SB templates of 
    M12 (magenta squares and asterisks); 6) four selected templates of
    B13 (yellow open triangles); 7) the templates of
    C14 (green crosses). 
    The data plotted in the figure
    include galaxies from Table \ref{tab:smglist}. The data from
    \citet{saintonge2013} are shown in gray squares, those from
  \citet{casey2012} and \citet{roseboom2012} are shown in gray
  crosses, and the data from other works are shown in gray dots.}
  \label{fig:comp_sed_data}
\end{figure*}

The $S_{250}/S_{350}$ versus $S_{500}/S_{350}$ values for
our high-z sample are plotted in Figure \ref{fig:comp_sed_data}. The
 sample is divided into four redshift bins:
 $z$ $\sim$ 3 ($2.5<z<3.5$), $z$ $\sim$ 4 ($3.5<z<4.5$), $z$ $\sim$ 5
($4.5<z<5.5$), and $z$ $\sim$ 6 ($5.5<z<6.5$). The colors given by MBB
models with different dust temperatures are also plotted. We
  assume optically thick MBB models, with $\nu_0=c/100\,\mu$m and
  $\beta=$1.8 (see also Section \ref{subsec:mbb}).

In all panels only three galaxies have $S_{500}/S_{350}$ colors larger
than $1.5$: the upper
left region of the color-color plot is avoided by all the
observations. The lack of galaxies with these colors may indicate
that at high-z most galaxies have intrinsic high dust temperatures.
Before reaching a conclusion we must check if the detection limits of
the observations could prevent us from selecting
cold dust objects: this is discussed in Section
\ref{sec:limits}. The sample of \citet{saintonge2013} (gray squares) seems to have
hotter dust temperatures than the other galaxies, which may be due to the
difference between UV- and IR-selected galaxies (Section
\ref{subsec:uvsel}).

In our high-z sample, the redshifts are measured from different
  wavelength. The reliability of optically measured redshifts
  given in \citet{casey2012} and \citet{roseboom2012} depends on
the cross-matching methods. These sources are shown as crosses in
Figure \ref{fig:comp_sed_data}. It seems at $z\sim 4$ these sources
have higher dust temperatures comparing to other sources, but removing
these sources does not affect our conclusions.

\subsection{Comparison of SED templates with data at $z>2.5$}
\label{subsec:data_sed}

\begin{table*}
\centering
\caption{Table of templates and the mean $\chi^2$ of each template to
  fit the data at different redshift bins.}
\begin{tabular}{c cc ccc}
\hline
templates & \multicolumn{5}{c}{$<\chi^2>$}\\
       & all & $2.5<z<3.5$ & $3.5<z<4.5$ & $4.5<z<5.5$ &
$5.5<z<6.5$ \\
\hline
\multicolumn{6}{l}{local libraries}\\
\hline
  CE01 $10^9L_{\odot}$ & 29.61 & 42.62 & 12.53 & 7.52 & 6.43 \\
  CE01 $10^{11}L_{\odot}$ & 7.09 & 9.40 & 4.24 & 1.93 & 3.97 \\
  CE01 $10^{13}L_{\odot}$ & 4.78 & 5.87 & 3.97 & 2.62 & 1.09 \\
  DH02 $\alpha=1.0$ & 6.01 & 7.04 & 5.33 & 4.62 & 1.32 \\
  DH02 $\alpha=1.5$ & 4.14 & 5.52 & 2.78 & 0.86 & 1.31 \\
  DH02 $\alpha=2.0$ & 19.77 & 28.21 & 8.83 & 4.82 & 5.08 \\
  DH02 $\alpha=2.5$ & 39.67 & 55.84 & 18.88 & 12.74 & 8.69 \\
  C14 & 32.56 & 44.22 & 17.28 & 13.87 & 10.33 \\
\hline
\multicolumn{6}{l}{high-z libraries}\\
\hline
  E11ms & 22.34 & 33.11 & 7.99 & 3.80 & 4.11 \\
  E11sb & 13.79 & 20.37 & 5.11 & 1.86 & 3.05 \\
  M12ms & 6.72 & 8.94 & 3.94 & 1.63 & 3.94 \\
  M12sb & 4.85 & 6.50 & 2.95 & 0.64 & 2.81 \\
  B13 SF\_1 & 32.56 & 40.59 & 21.70 & 21.57 & 16.06 \\
  B13 SF\_2 & 9.78 & 13.59 & 4.61 & 2.13 & 4.98 \\
  B13 Obs\_SF & 6.87 & 7.80 & 6.37 & 6.17 & 1.59 \\
  B13 MIRex\_SF & 9.03 & 13.35 & 3.54 & 1.00 & 1.66 \\
\hline
\end{tabular}
\label{tab:sedlist}
\end{table*}

To quantify the coherence between templates and data, we fit
each SED template to all the high-z galaxies and calculate the
$\chi^2$ value for each galaxy. The mean value of the $\chi^2$
($<\chi^2>$) is used as an indicator of the average quality of the
fits with a given template. During the fit, we assume a confusion
noise of $6$ mJy for the galaxies with no confusion noise
estimation. The results are
shown in Table \ref{tab:sedlist}. The colors corresponding to the
templates used in this comparison (both local and for high-z objects)
are also plotted in Figure \ref{fig:comp_sed_data}.

We briefly discuss the case of the local templates from CE01, DH02, and
C14: we do not expect a good agreement since the local calibrations
are known to be invalid at high-z \citep{casey2014,lutz2014}.
Indeed, Figure \ref{fig:comp_sed_data} shows that the fixed template
of C14 is too cold to describe the high-z galaxies.
The CE01 and DH02 templates can be adjusted to fit the emission of
high-z populations. The
template of CE01 corresponding to  $L_{\rm IR}$ to $\sim
10^{11}L_{\odot}$ can reproduce the SPIRE observations of our high-z sample.
However, the local calibration of these templates cannot be used to give correct
$L_{\rm IR}$ measurements. For the DH02 templates, the lowest value of
$<\chi^2>$ is obtained for $\alpha$
$=$ $1.5$, a value lower than the average one found for galaxies at
$z \le 2$ ($\alpha\ =\ 2$, \citealt{buat2011,buat2012}).

The high-z libraries include the templates of E11, M12, and B13. The MS
template of E11 predicts no evolution with redshift of
the SED shape for MS galaxies. With this template $<\chi^2>=22.34$
indicating the poor quality of the fit. Figure \ref{fig:comp_sed_data}
shows that the E11 MS template produces colors colder than
most observations at $z>3$, implying that the template is
biased to galaxies with cold dust temperatures. The SB template of E11
gives a better fit to the
observation data, but is still colder than most observations and with a high value of $<\chi^2>$ ($<\chi^2>=13.79$).

Assuming a mild evolution of the MS SEDs, the M12 templates show a
better consistency with our high-z sample ($<\chi^2>=6.72$). Here we use
the U8 template, which is assigned to $z \sim 3$ by M12, assuming no
further evolution for redshifts $>$ $3$. Figure
\ref{fig:comp_sed_data} shows that the U8 template is able to
reproduce the average observed colors at $z \sim 4$, $5,$ and $6$,
in agreement with M12's assumption that the SEDs evolve very
little at $z>3$ because of the flattening of the evolution of the
sSFR. The SB template of M12 gives smaller
values of $<\chi^2>$ (4.85) than the MS template. M12 used G20
to derive their SB template. Removing this galaxy from the high-z
sample, the $<\chi^2>$ becomes 7.35 for the MS template, and 5.22 for
the SB template. The SB template still fits slightly better than the MS
template, implying most of our high-z galaxies more likely
belong to SB population.

The $<\chi^2>$ values obtained with the B13 models are intermediate.
The colors of the B13 template for
star-forming galaxies in GOODS field (sf\_glx\_1) extend to colder
regions than observations. The bestfit template in the B13 library is
the template for obscured star-forming galaxies (Obs\_SF\_glx), which
gives the $<\chi^2>$ of 6.87.

\begin{figure*}
\centering
\includegraphics[width=0.9\linewidth]{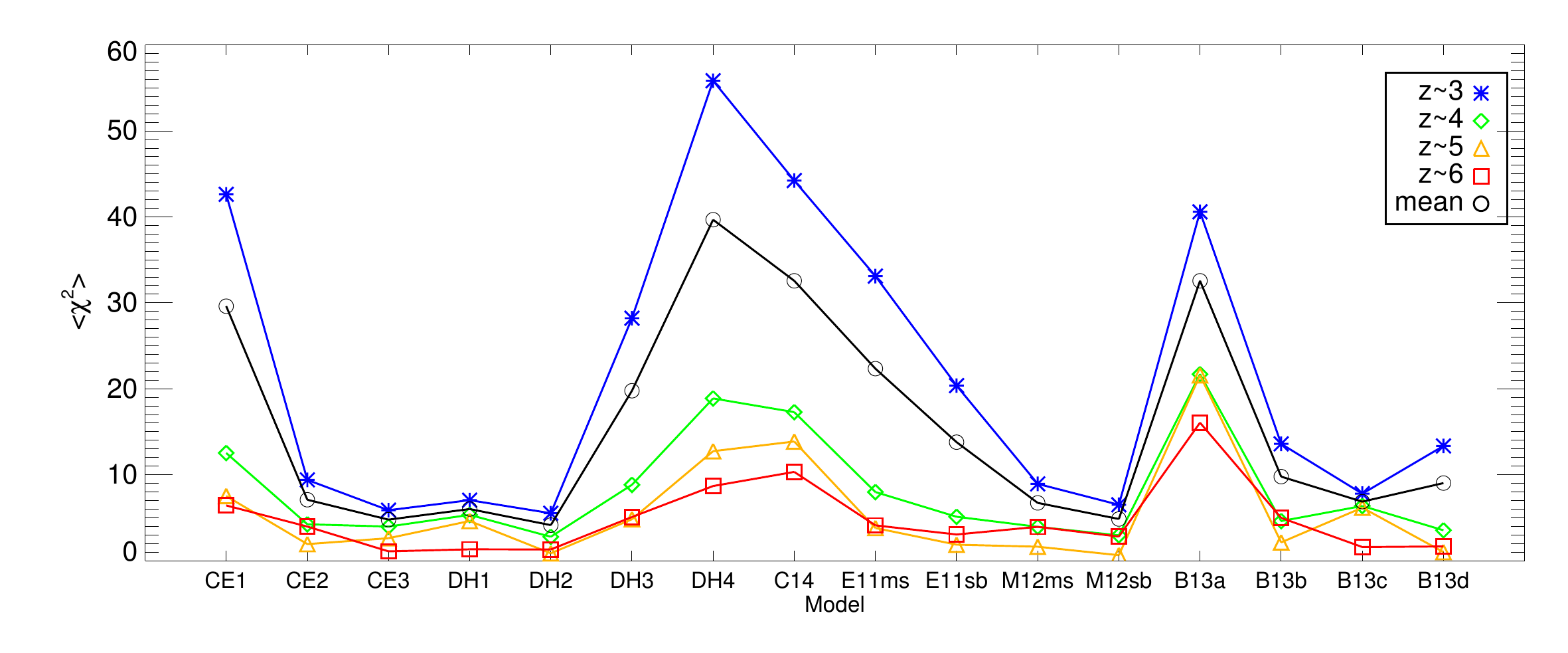}
\caption{The mean $\chi^2$ for templates
  listed in \ref{tab:sedlist} at $z\sim3$ (blue line with asterisks),
  $z\sim4$ (green line with diamonds), $z\sim5$ (yellow line with
  triangles), and $z\sim6$ (red line with squares). The mean $\chi^2$
  for all the galaxies are plotted in black line with circles. In the figure, we use CE1, CE2, and CE3 to represent CE01 templates with $10^9L_{\odot}$, $10^{11}L_{\odot}$ , and $10^{13}L_{\odot}$, respectively. We use DH1, DH2, DH3, and DH4 to represent DH02 templates with $\alpha=1.0$, $1.5$, $2.0,$ and $2.5$, respectively. We use B13a, B13b, B13c, and B13d to represent  the SF\_glx\_1, SF\_glx\_2, Obs\_SF\_glx, MIRex\_SF\_glx templates in the B13 library, respectively.}
\label{fig:chi2}
\end{figure*}

We also calculate the $<\chi^2>$ in different redshift
  bins. The result for each template is shown in Table
  \ref{tab:sedlist} and Figure
  \ref{fig:chi2}. The $<\chi^2>$ for
each template at $z\sim6$ becomes similar, which means
that the goodness of the fit becomes less dependent on the templates
because of the large uncertainties of the data. The large uncertainties of
the data at $z\sim 6$ causes the $<\chi^2>$ value of the fitting at
$z\sim6$ to be smaller than at $z\sim3$. The the $<\chi^2>$ value
are smaller at $z\sim6$ also because there
are only four sources at $z\sim6$ and thus less scatter between the sources.

CE01 template with $L_{\rm IR}$ to $\sim
10^{11}L_{\odot}$, DH02 template with $\alpha=1.5$ and M12
  template fit the data well at each redshift bin. It is possible that galaxies with these kinds of templates are major
populations at high-z, but the better fitting can also be due to
the data not being sufficient to constrain the SED. The local templates
CE01 and DH02 give similar or better $<\chi^2>$ than high-z
templates, probably because they have flexible parameters ($\alpha$ in
DH02 and $L_{\rm IR}$ in CE01).

  Since the SPIRE beam is larger at longer wavelengths, the
  blending affects $500\mu$m fluxes more than $350\mu$m and $250\mu$m
  fluxes. Therefore, the observed SEDs may be overestimated at longer
  wavelength. If the effect is non-negligible, the intrinsic colors of
  the galaxies is bluer than those plotted in Figure
  \ref{fig:comp_sed_data}. The templates of CE01 with
  $L_{IR}>10^{11}L_{\odot}$, DH02 with $\alpha<1.5$, the SB
  template of M12, and the Obs\_SF\_glx template of B13 are still preferential.

\subsection{UV-selected galaxies on the SPIRE color-color diagram}
\label{subsec:uvsel}

\citet{saintonge2013} have shown that their galaxies are
similar to UV-selected LBGs, with a lower $L_{\rm IR}$
comparing with the IR/submillimeter-selected galaxies. Figure
\ref{fig:comp_sed_data} shows that the galaxies from
\citet{saintonge2013} are located in hotter regions than the other
IR/submillimeter-selected galaxies (based on the MBB models), probably
indicating hotter dust temperatures for these UV-selected galaxies.

\begin{figure}
\centering
\includegraphics[width=\linewidth]{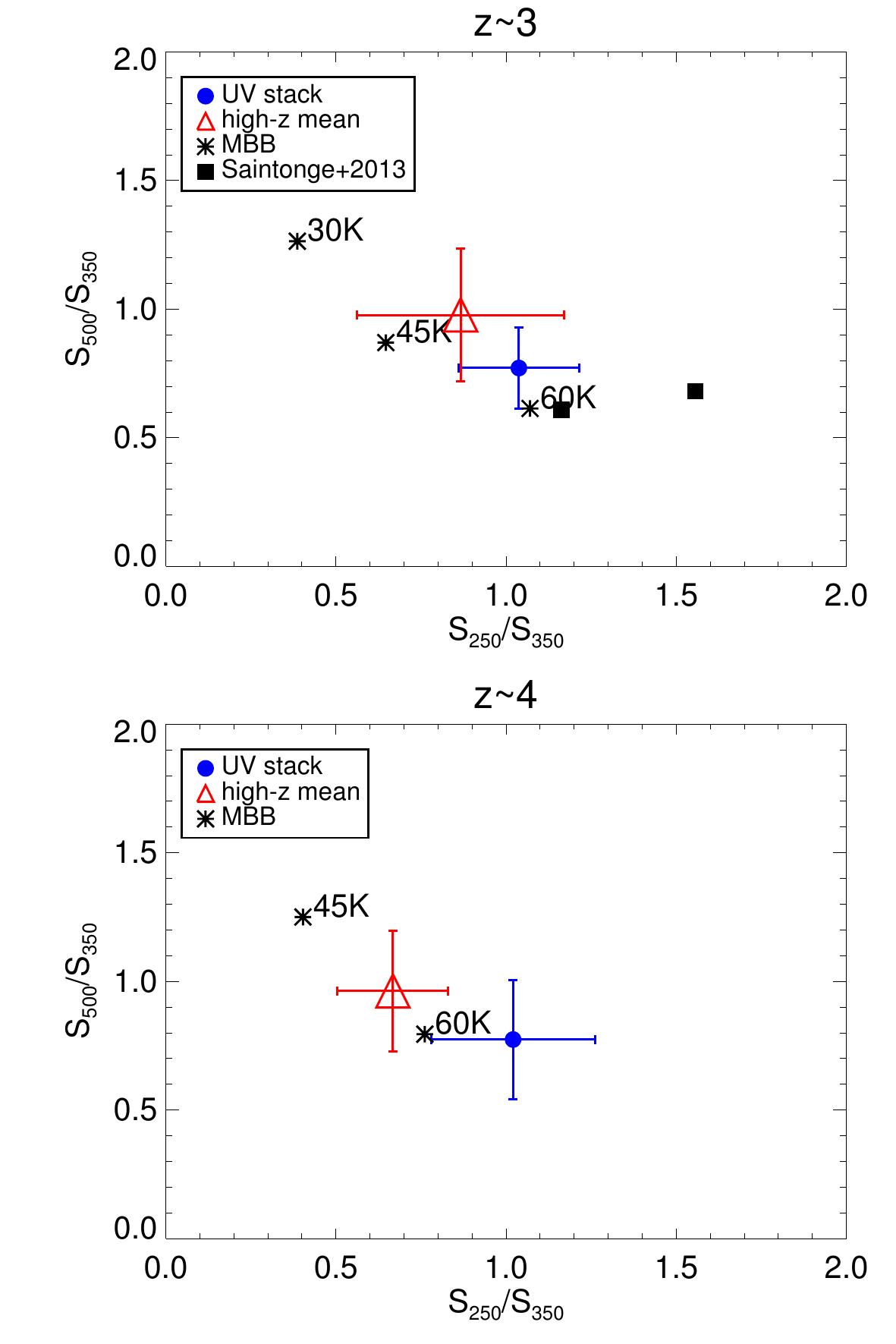}
\caption{Comparison of SPIRE colors for UV-selected galaxies
  from \citet{heinis2014} (blue) and our high-z sample (red) at
  $z\sim3$ (upper panel) and $z\sim4$ (lower panel). In each panel,
  the blue dots show the colors derived from the stacked UV sample
  from \citet{heinis2014}.The red triangle indicates the
  average colors of our high-z sample. The error bar indicates the
  standard deviation. Filled black squares
  present colors for the galaxies from \citet{saintonge2013}. The
  asterisks indicate the colors given by MBB models with
  dust temperatures of 30K, 45K, and 60K.}
\label{fig:plot_seb}
\end{figure}

From a stacking analysis in the COSMOS field, \citet{heinis2014}
measured average IR fluxes for samples of UV-selected
galaxies at $z\sim3$ (20849 galaxies) and $\sim4$ (6794 galaxies) (see
  also Section \ref{subsec:compsamp} for the details of this
  sample). Comparing these samples with our high-z
sample and the sample of \citet{saintonge2013} in Figure
\ref{fig:plot_seb}, we find that, as with the sample of galaxies in
\citet{saintonge2013}, the SPIRE colors of these
UV-selected samples are slightly hotter than the average colors of
our high-z sample of IR/submillimeter-selected galaxies.

A high dust temperature in a UV-bright
galaxy can be caused by a strong UV radiation
intensity, which can heat the dust grains in the
star-forming regions to high temperatures. The high dust temperatures
and low-IR luminosities of the UV-selected sample as compared to
those of the IR/submillimeter-selected galaxies indicate that the dust mass of
the UV-selected galaxy is also low
($M_{d}\propto L_{\rm IR}T_{d}^{-4+\beta}$)
\citep{casey2014}. In IR/submillimeter-selected galaxies, the radiation field may
evolve less than estimated by previous studies, which use UV-selected
galaxies to probe their evolution \citep[e.g.,][]{saintonge2013}.
A detailed comparison of high-z UV-selected
and IR/submillimeter-selected galaxies is beyond the scope of this paper and will be
addressed in a future work.

In summary, from the comparison between the data and
templates we find the following: 1) most galaxies  of our
sample (except 3 of them)  have $S_{500}/S_{250}\ <\ 1.5$,
inconsistent with  MBB models with cold dust temperatures or local
mean SED templates (C14);  2) to fit  the observations at
high-z,  the parameter $L_{\rm IR}$ of the CE01 templates should be adjusted
to $\sim 10^{11}L_{\odot}$, and the parameter $\alpha$ of the DH02
template should be adjusted to $\sim 1.5$; 3) the E11 MS template
corresponds to colder colors than the average of the observations; 4) the
MS template of M12 can be used for high-z galaxies ($z\ >\
2.5$) to reproduce the average colors given by the observations;  and 5)
the colors of UV-selected galaxies correspond to hotter dust temperatures than
those of submillimeter/IR-selected galaxies.

\section{Discussion}
\label{sec:discussion}
\subsection{Detection limits}
\label{sec:limits}
In Section \ref{sec:res} we showed that there are no observations
within the upper left regions of the SPIRE color-color diagram, which
is inconsistent with a number of SED libraries. This can be due to
the intrinsic high dust temperature of high-z sources and/or a selection
effect due to the detection limits. In this section, we check whether
the detection limits prevent us from observing objects in the upper left
regions of the SPIRE color-color diagram.

\begin{figure}
\centering
\includegraphics[width=\linewidth]{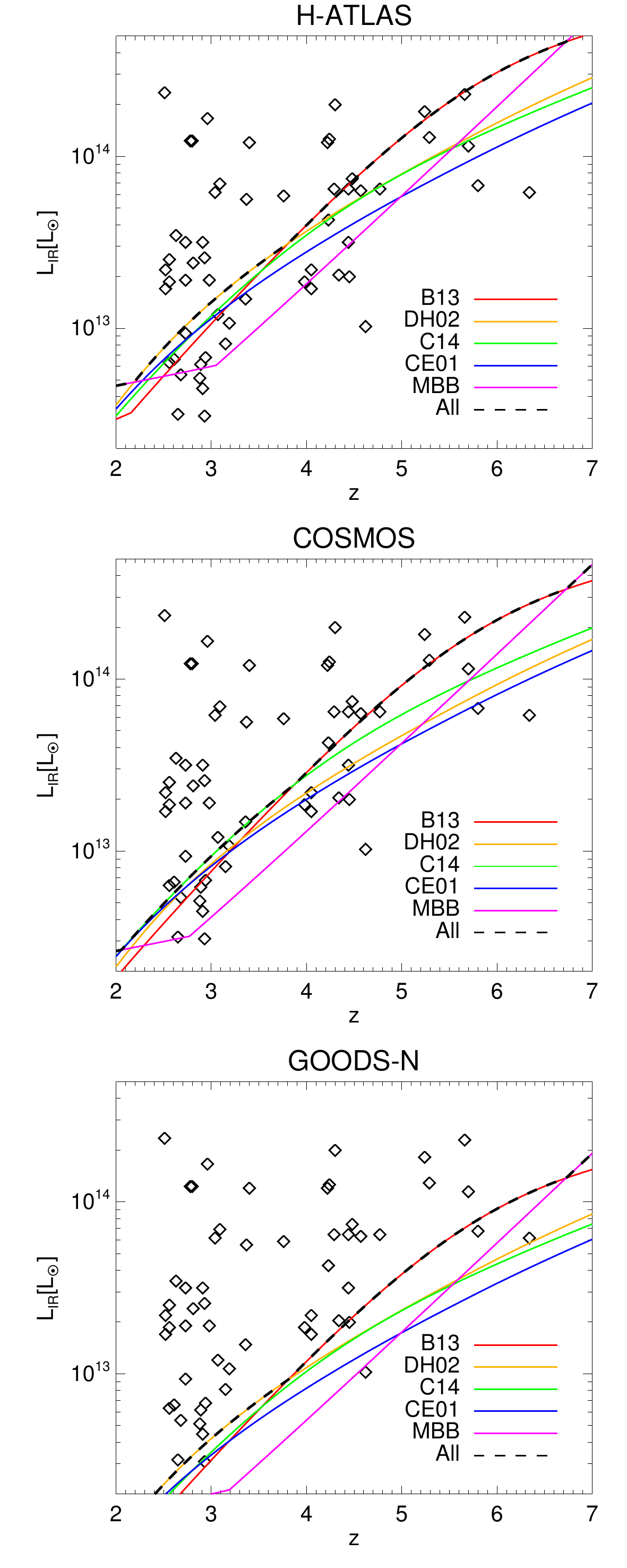}
\caption{Completeness limits $L_{\rm IR}$ for detecting all SED
  shapes. The upper, middle, and bottom panels show the complete
  limits for H-ATLAS, COSMOS, and GOODS-N surveys, respectively. In
  each panel, the galaxies of our high-z sample are plotted as
  diamonds ($L_{\rm IR}$ from Table \ref{tab:smglist}), and the
  $L_{\rm IR}$ for each SED template is plotted in colored line (red:
  B13; yellow: DH02
  with $\alpha=2.5$; green: C14; blue: CE01 with $L_{\rm
    IR}=10^9L_{\odot}$; magenta: MBB with $T=40K$). The black line is
  the maximum of the colored lines, above which the observation should
  not be biased against these SEDs.}
\label{fig:complete}
\end{figure}

First, it is possible that the high-z sample is biased against
galaxies with certain types of SEDs, because a galaxy only appears on the
color-color diagram  if it is detected in the three SPIRE bands
(color selection effect). Therefore, we check whether the
 detection limits of our high-z sample have an impact on the type of
  SEDs we can probe. We examined
  five SEDs that are less favored by the observations. They are MBB
  models with
  $T=40$K, the DH02 template with $\alpha=2.5$, the CE01 template with
  $L\sim10^{9}L_{\odot}$, and the SF\_glx\_1 template of B13 and C14 template. The minimum $L_{\rm IR}$ required to observe them are
  explored by defining the complete $L_{\rm IR}$-$z$ parameter space
  for the H-ATLAS, COSMOS, and GOODS-N surveys.
  The curve $L_{\rm IR}^{\rm complete}(z)$ is determined as follows:
  \begin{enumerate}

\item Calculate the minimum detectable IR luminosity at $z$ for each
  SED ($L_{\rm IR}^{\rm lim}(S(\lambda),z)$) from the SED template
  $S(\lambda)$ and the detection limit $S_{\lambda}^{\rm lim}$ (Table
  \ref{tab:depths}) for each survey. The limit for each SED template
  is shown as colored curves in Figure \ref{fig:complete}.
\item Take the maximum of $L_{\rm IR}^{\rm lim}(S(\lambda),z)$ at each
  $z$. The result $L_{\rm IR}^{\rm complete}(z)$ is the luminosity
  required to detect a source having any of the SEDs listed in Table
  \ref{tab:depths} in the three SPIRE bands (black curve in
  Figure \ref{fig:complete}).
\end{enumerate}

The region above the black curve in each panel shows the $L_{\rm
  IR}$-$z$ range where sources with these five SEDs are all detectable
(complete for all SED shapes).

We find that 27 of the 57 galaxies in our sample are within the
H-ATLAS completeness limit. For the GOODS detection limits, the number
of such galaxies increases to 52. In the worst case of the H-ATLAS
detection limits, we have 19 galaxies at $z\sim3$ and 5 galaxies at
$z\sim4$ within the completeness limits, whereas $z\sim5$ and $6$
samples are not complete. For the deepest field, the nondetection of
galaxies with these templates intrinsically indicates  there are fewer  or no objects with
 such SED shapes rather than the selection effect at $z\sim3$ or $4$,
implying that the dust temperature is intrinsically high for high-z
galaxies. In fact, if a galaxy with $T=30$K is above the GOODS
detection limit, the dust mass required is larger than
$10^{10}M_{\odot}$ at $z>4$. It is not surprising that these kinds of galaxies are rare. 

\begin{table}
  \centering
  \caption{The 3$\sigma$ depths of H-ATLAS, COSMOS, and GOODS,
    including confusion noise.}
  \begin{adjustbox}{max width=\linewidth} 
  \begin{tabular}{c c c c c}
    \hline
    Survey  & \multicolumn{3}{c}{Depths[mJy]}           & reference\\
    & 250 $\mu$m & 350 $\mu$m & 500 $\mu$m &          \\
    \hline
    H-ATLAS & 18.0 & 21.0 & 27.0 & \citet{smith2011}\\
    COSMOS  & 15.2  &  17.0 & 19.2 & \citet{oliver2012}\\
    GOODS   & 5.7  &  7.2 &  9.0 & \citet{elbaz2011}\\
    
    \hline
  \end{tabular}
  \end{adjustbox}
  \label{tab:depths}
\end{table}

Another question is whether galaxies that are fainter than the completeness
limits are colder. Examining the properties of our UV-selected sample,
we found that the FIR faint objects at z $\sim$ 4 are not located in the
upper left region, suggesting that they on average have high dust
temperature. Therefore, the UV-selected galaxies with cold dust
  temperatures may exist, but should be much rarer than
  galaxies with high dust temperatures. One might argue that galaxies
may exist  with
no detectable\ UV nor FIR, however, the lensed galaxies in our
sample are intrinsically faint, and there is no galaxy located in the
upper left region of the color-color diagram. Therefore, it is
  likely that the cold dust galaxies are intrinsically few at
  high-z. The high dust temperature of high-z
galaxies may be due to the strong star formation activity in these
galaxies.

However, the argument is not conclusive because our sample still
  has some potential selection biases toward
  luminous galaxies. First, our requirements that the
galaxies have spectroscopic redshifts can bias the sample to be
brighter than the formal detection limits. This effect may not be
significant since similar results
can be found for the sample of \citet{swinbank2014}, who used the ALMA
LESS sample with photometric redshifts $2<z<5$:
few galaxies are found with $S_{500}/S_{350}\ >\ 1.5$ (see their
Figure 5). Second, although we can explore galaxies fainter than the
detection limits taking advantage of lensing magnification, the
faintest galaxies in our sample have $L_{\rm
  IR}\sim10^{11}L_{\odot}$. It is possible that cold dust exists in
galaxies fainter than $\sim10^{11}L_{\odot}$.

\subsection{Sampling high-z galaxies using the SPIRE color-color
  diagram}
\label{sec:photoz}

\begin{figure}
  \centering
  \includegraphics[width=\linewidth]{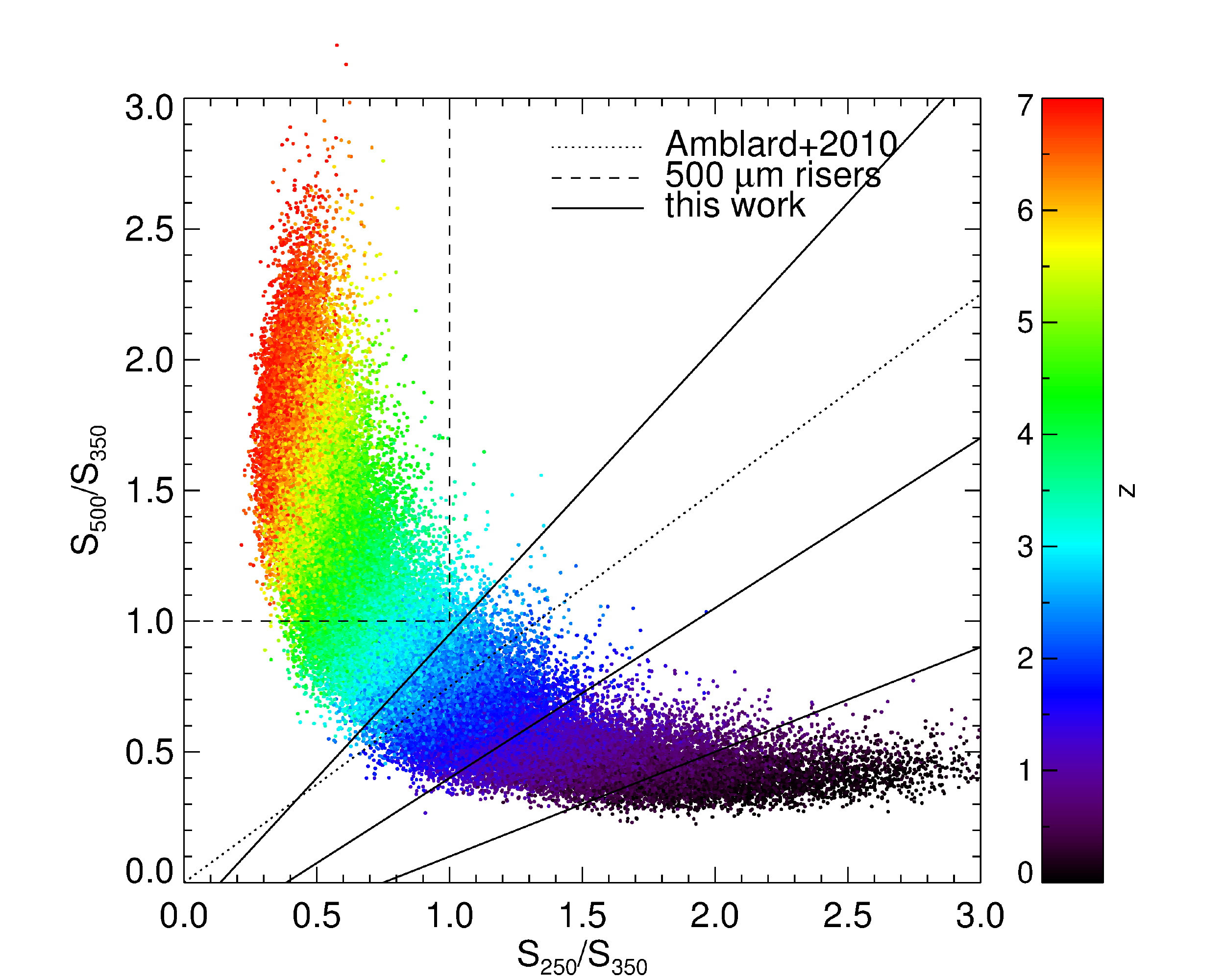}
  \caption{Color-color diagram of the 70,000 mock galaxies
    generated from M12 templates. The color indicates the redshift of
    the mock galaxy. The dotted line shows the color cut given by
    \citet{amblard2010}. Galaxies above this line are expected to
      have $z\sim2.6$ and those under it are expected to have
      $z\sim1.8$. The dashed line shows the selection of $500$
    $\mu$m risers. The upper left region separated by this line
    selects galaxies with $z>4$. The solid lines are our color cut method according to
    the distribution of the mock galaxies. The solid lines
      divided the diagram into four parts. From lower right to upper left
      these parts select galaxies at $z<0.5$,
      $0.5<z<1.5$, $1.5<z<2.5$ and $z>2.5$.}
  \label{fig:photoz}
\end{figure}

At high redshifts, since the available spectroscopic information is
extremely limited, it would be useful to be able to select high-z
candidates from photometric data. The SPIRE colors have been used for
selecting high-z candidates in several studies. For example, some
studies select 500 $\mu$m risers with $S_{500}>S_{350}>S_{250}$
\citep[e.g.,][]{riechers2013,dowell2014} as high-z candidates. Galaxies
with $S_{500}>S_{350}>S_{250}$ are likely to be at redshift $z>4$
according to the shape of the IR SED. The galaxy HFLS3 at $z\sim6$ was
found through this method. Another method is to model the redshift
distributions on the SPIRE color-color diagram and compare with
observations \citep{amblard2010, roseboom2012}. From the redshift
distributions, \citet{amblard2010} also define a color cut with
$S_{500}/S_{250} >0.75$ and $<0.75$, which divides their H-ATLAS
galaxy sample into $z \sim 2.6$ and $z \sim 1.8$ populations.

However, all these methods are based on MBB models, where the redshift
evolution of the SEDs is not included. Therefore, more sophisticated
SED templates should be used to investigate the correlation between
redshift and colors. We showed in Section \ref{sec:res} that the M12
templates can reproduce the SPIRE colors that are consistent with the
high-z observations. The M12 library also gives the
evolved SEDs for $z<3$. Therefore, in this section we use the M12
library to re-investigate these methods.

Using the MS templates of M12, we construct 70,000 mock galaxies with
$z \in [0.0,7.0]$. We added an extra Gaussian standard deviation of
$10$\% to the fluxes and then
calculate the colors of these mock galaxies.
In previous studies, $10$\% is commonly used 
\citep[e.g.,][]{amblard2010,robson2014}. Using reasonably larger values 
($20$\%) does not affect our conclusion. 

We overplotted the selection methods of 500 $\mu$m riser (dashed
line) and \citeauthor{amblard2010} (2010; dotted line) in Figure
\ref{fig:photoz}. Both methods are based on MBB models. It seems 
that the 500 $\mu$m riser method selects a
very pure high-z sample. The color cut
defined by \citet{amblard2010} is able to divide the mock galaxy
sample into high-z ($z \sim 2.6$) and low-z ($z \sim 1.8$)
populations, whereas the redshift range
in each population is quite large.

As shown in Figure \ref{fig:photoz} $z>3$ galaxies are located in 
a smaller area on the color-color diagram predicted by the M12 library 
comparing with that predicted by MBB models (see Figure 1 in 
\citealt{amblard2010}). Based on the distribution predicted by the M12 
library, we redefine the color cuts as follows:

\begin{eqnarray}
  \label{equ:r0}
  \frac{S_{500}}{S_{350}} < 0.4\,\frac{S_{250}}{S_{350}}-0.3. \\
  0.4\,\frac{S_{250}}{S_{350}}-0.3 \leq  \frac{S_{500}}{S_{350}} <
  0.65\,\frac{S_{250}}{S_{350}}-0.25\\
  0.65\,\frac{S_{250}}{S_{350}}-0.25 \leq \frac{S_{500}}{S_{350}} <
  1.1\,\frac{S_{250}}{S_{350}}-0.15\\
  \frac{S_{500}}{S_{350}} \geq 1.1\,\frac{S_{250}}{S_{350}}-0.15
  \label{equ:r3}
.\end{eqnarray}
Equations \ref{equ:r0} to \ref{equ:r3} define four regions
R0($z<0.5$), R1($0.5<z<1.5$), R2($1.5<z<2.5$) and R3($z>2.5$) in Figure
\ref{fig:photoz}, separated by the three solid lines.

\begin{figure}
\includegraphics[width=\linewidth]{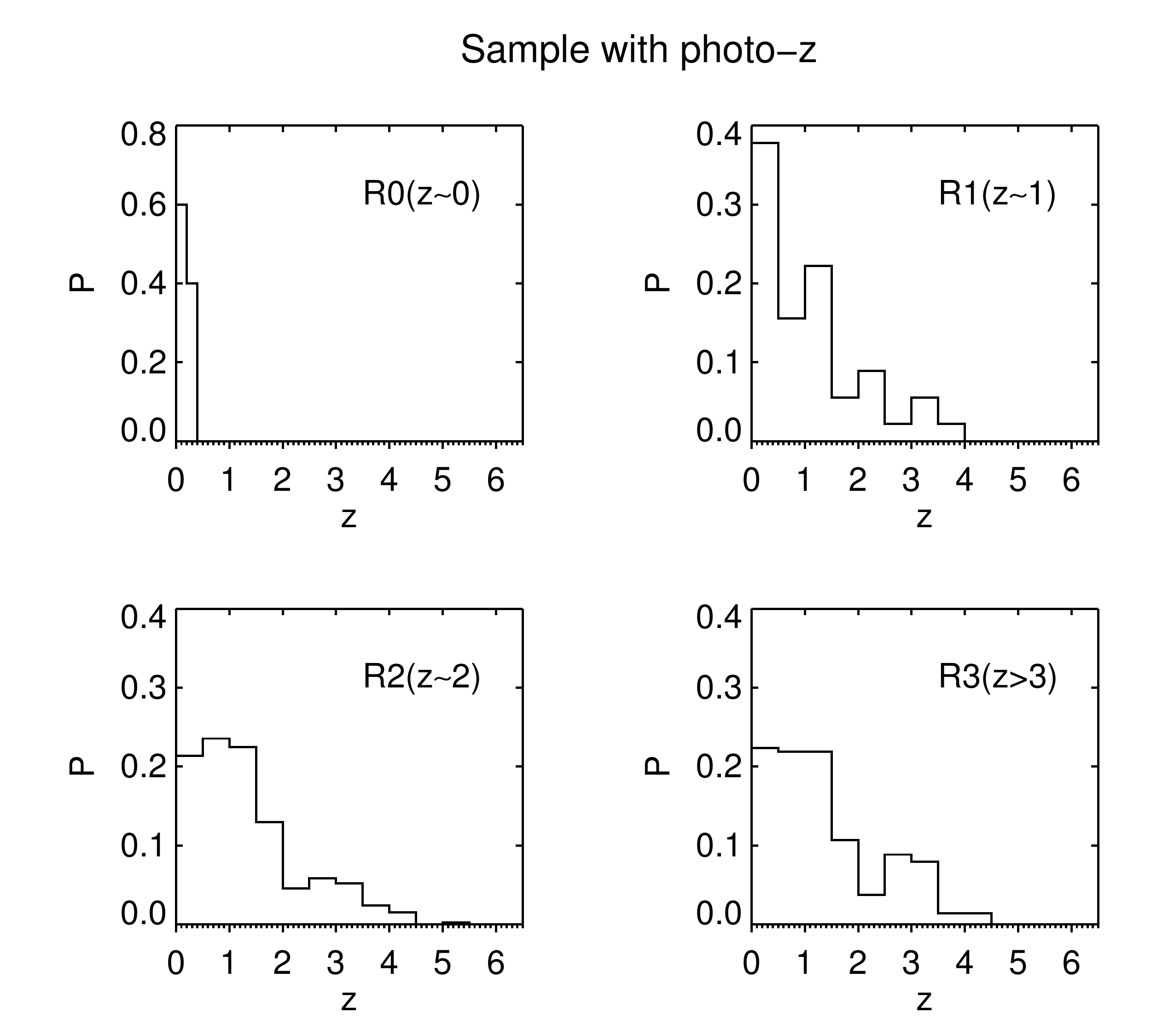}
\caption{Redshift distributions of the four subsamples of the 783
  galaxies divided using the color-cut method defined in this work
  (Equations \ref{equ:r0} to \ref{equ:r3}).}
\label{fig:photoz_distri}
\end{figure}

\begin{figure}
\includegraphics[width=\linewidth]{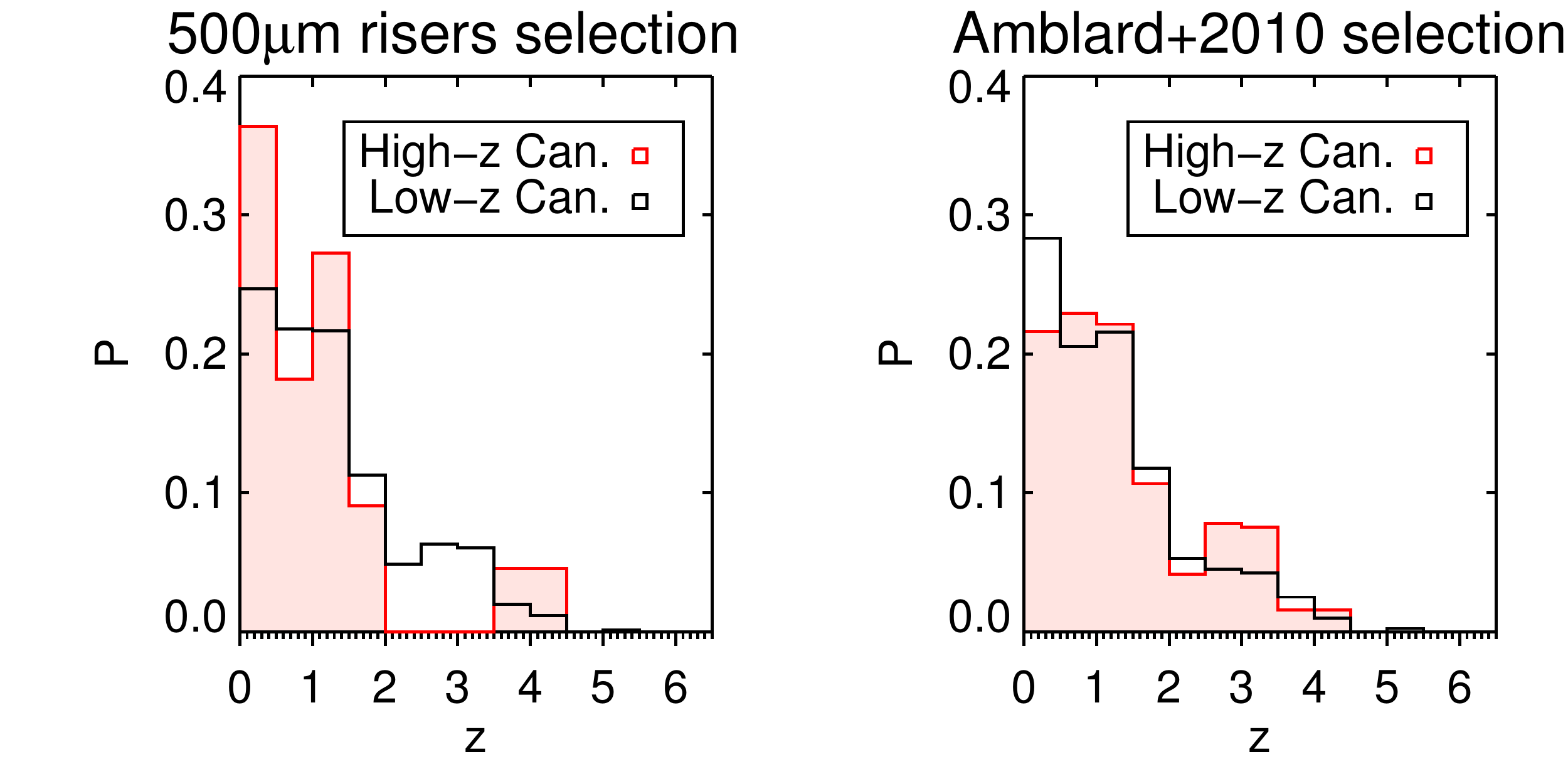}
\caption{Photometric redshift probability distributions of a
  sample of 250 $\mu$m selected galaxies. Left panel: The filled
  histogram shows the redshift distribution for the subsample of 500
  $\mu$m risers ($S_{500}>S_{350}>S_{250}$) and the open histograms
  show the redshift distribution for the remaining galaxies in the
  sample. Right panel: The filled histograms show the redshift
  distribution of the high-z subsample ($z\sim2.6$) selected using
  \citet{amblard2010}'s color cut ($S_{500}/S_{250}>0.75$), and the
  open histograms show the redshift distribution of the low-z
  subsample ($z\sim1.8$) selected using $S_{500}/S_{250}<0.75$.}
\label{fig:photoz_other}
\end{figure}

In the case of our spectroscopically confirmed high-z sample (57
galaxies at $z>2.5$), 40 galaxies are in R3
region. \citet{amblard2010}'s high-z color cut selects 42 of these
galaxies. The method of 500 $\mu$m risers selects 23 of these
galaxies. The completeness of the 500 $\mu$m riser selection is
$40$\%, worse than both our method work and \citet{amblard2010}. Since
the 500 $\mu$m riser selection method is designed for selecting $z>4$
galaxies, we also check the selection completeness for $z>4$ galaxies
in our sample and found the completeness is only $\text{approximately }66\%$ (14 of 21).

To further test the three methods, we use the $500$ $\mu$m selected
sample with photometric redshift information in \citeauthor{rr2014} (2014; 783
galaxies, see Section \ref{subsec:compsamp} for details of this
sample). The large size of this sample is necessary to test the
color-cut methods because these methods are designed for  statistical
use. Figure \ref{fig:photoz_distri} shows the redshift distributions
of the four subsamples of the 783 galaxies divided using the color-cut
method defined in this work. The redshift distributions of subsamples
divided using the methods of 500 $\mu$m risers and \citet{amblard2010}
are shown in Figure \ref{fig:photoz_other}. The mean median redshift,
the standard deviation, and the contamination from other redshifts for
each subsample are listed in Table \ref{tab:statisticp}.

\begin{table}
  \centering
  \caption{Statistics of the redshift distribution (photometric
  $z$) in each region on the color-color diagram divided using
  Equations \ref{equ:r0} to \ref{equ:r3} and other methods.}
  
  \begin{adjustbox}{max width=\linewidth}                               
  \begin{tabular}{cccccc}
    \hline
    &  N &Mean       & Median   &   STD & contamination\\
    \hline
    All & 783 & 1.31 & 1.06 & 1.00 \\
    R0 ($z<0.5$)  & 15 & 0.18 & 0.16 & 0.09 & 0\%\\
    R1 ($0.5<z<1.5$) & 90 & 1.09 & 0.96 & 0.94 & 62\%\\
    R2 ($1.5<z<2.5$) & 463 & 1.35 & 1.08 & 0.99 & 83\%\\
    R3 ($z>2.5$) & 215 & 1.39 & 1.13 & 1.03 & 80\%\\
    Amblard+($z>2$) & 384 & 1.39 & 1.12 & 1.02 & 77\%\\
    Amblard+($z<2$) & 399 & 1.23 & 1.02 & 0.98 & 18\%\\
    500 $\mu$m risers ($z>4$)  & 22 & 1.13 & 0.96 & 1.09 & 95\%\\
  \hline
  \end{tabular}
  \end{adjustbox}
  \label{tab:statisticp}
\end{table}

From the statistics, we find that at $z<1.5$ the color cuts
defined in this
work can separate galaxies into subsamples with different
mean redshifts, and the Student's t-test shows that the difference
between these subsamples is significant (the index $\alpha<0.05$). The
mean redshift of galaxies is 0.18 in R0 region, and 0.96 in
R1 region, consistent with our definition of these regions using mock
galaxies. However, at $z>1.5$ the mean redshifts in R2 and R3
regions are lower than expected. The high- and low-redshift subsamples
given by the method of \citet{amblard2010} have mean redshifts of 1.43
and 1.23, respectively. The Student's t-test shows that the difference is
also significant. However, the mean redshifts are different from
\citet{amblard2010} values for H-ATLAS galaxies (2.6 for the
high-z population and 1.8 for the low-z population). The method of 500
$\mu$m risers selects a high-z sample with mean redshift of 1.53,
significantly lower than the expectation for this method. The contamination becomes higher than 50\% at $z>2$ for all the
methods, and the dispersion of redshifts in all the $z>0.5$
subsamples in Table \ref{tab:statisticp} is quite large
($\sim1$). 

The contamination is partly because of the sample selection. The
requirement of association with $24\mu$m source and an entry in the SWIRE Photometric Redshift Catalog biases the sample
against sources with $S_{500}/S_{24}>200$ as well as sources with
$z>1.5$ \citep{rr2008,rr2014}. Our requirement of detection at
$250\mu$m further biases this sample to lower redshifts. The lack of
$z>1.5$ sources in this sample therefore causes poor statistics at
$z>1.5$. However, assuming galaxies without $24\mu$m or $250\mu$m
detection are all perfectly divided by the color-cut methods, we can
estimate a minimum contamination rate, which is $53\%$ for R3 ($z>2.5$
cut), $53\%$ for \citet{amblard2010}'s $z>2$ cut, and $50\%$ for
$500\mu$m risers cut. The minimum contamination is still high for high-redshift color cut. The error of photometric redshift and the
cross-matching accuracy is
also relevant to the contamination. The photometric
redshift error for this sample is about $4\%$ in $(1+z)$ 
\citep{rr2014}, which increases with the redshift. At higher
redshift, the association of submillimeter sources with $24\mu$m
sources also becomes less reliable.
Apart from the above causes, there is also intrinsic dispersion caused by
the variation of galaxy properties because we are considering an
average SED in a given redshift bin when designing the method.

From the discussion above, it seems that although the three methods
can select samples with different mean redshifts, the resulting
samples have either a large dispersion in $z$ or miss a considerable
number of galaxies ($\sim33$\%) at high redshift. Therefore, it is
difficult to properly sample galaxy
populations at different redshifts using only the three SPIRE bands.

\section{Conclusion}
\label{sec:conc}

We collected a sample of 57 galaxies with reliable redshifts and SPIRE
flux measurements at $z > 2.5$, and compare the SPIRE colors of this
sample with those derived from different SED templates at different
redshifts. The templates are taken from the pre-{\sl Herschel} and
post-{\sl Herschel} SED libraries, including local templates of CE01, DH02,
and C14, and high-z templates of E11, M12, and B13.

From the SPIRE color-color diagram, we find that $\sim 95\%$ of the
galaxies at $z>2.5$ have a SPIRE color $S_{500}/S_{350}<1.5$,
implicating that the dust temperatures in these galaxies are
high. Therefore, local SED templates may not be suitable for these
galaxies due to dust temperature evolution. From the comparison of the
high-z data with the local templates, we find that the SPIRE colors
given by the C14 template, which is based on local galaxies that are too
cold to describe the average colors of our high-z sample. The local
calibrations for CE01 and DH02 libraries are also not able to describe
the average SPIRE colors for high-z galaxies. However, both the CE01
and DH02 templates can fit the SPIRE colors better when the parameter
$L_{\rm IR}$ of the CE01 templates is adjusted to $\sim
\ 10^{11}L_{\odot}$ or the parameter $\alpha$ of the DH02 templates
fixed to $1.5$. Among the high-z templates, the templates of M12 give
the most satisfying fit, implying that their assumption of the SED
evolution is suitable for the high redshift.

Comparing with UV-selected samples at $z=3$ and $4$, we find that
  the UV-selected sample have even higher dust temperatures than our
  sample. We discussed the possible bias of the samples due to the
  detection limits, and find that the higher than local dust
  temperature for high-z galaxies is unlikely to be due to
  observational biases.

The efficiency of selecting high-z galaxies using the SPIRE
color-color diagram is also discussed. We defined color cuts according
to the MS template of M12 to sample galaxies at different redshifts,
and compare the method with the method using 500 $\mu$m risers and the
color cut defined by \citet{amblard2010}. We have shown that these
methods can divide our $250$ $\mu$m selected sample into subsamples
with different mean redshifts., however, the dispersion in each
subsample is quite large. Additional information is needed for better
sampling.

\begin{acknowledgements}
We thank an anonymous referee for his/her very useful comments and
suggestions. YFT is a LAMOST fellow and this work is supported by NSFC with
No. 11303070 (PI : YFT), No. 11433003 (PI Shu Chenggang), and
No.11173044 (PI HJL). This work is also supported by
the Strategic Priority Research Program ``The Emergence of Cosmological
Structures'' of the Chinese Academy of Sciences (CAS; grants
XDB09010100 and XDB09030200). SSY is supported by 973 Program
CB8845705, SZY is supported by the NSFC Major Project No. 11390373. LC
acknowledges financial support from  the {\sc thales} project 
383549 that is jointly funded by the European Union and the Greek 
Government in the framework  of the programme ``Education and lifelong 
learning''.

SPIRE has been developed by a consortium of institutes led by Cardiff
Univ. (UK) and including Univ. Lethbridge (Canada); NAOC (China); CEA,
LAM (France); IFSI, Univ. Padua (Italy); IAC (Spain); Stockholm
Observatory (Sweden); Imperial College London, RAL, UCL-MSSL, UKATC,
Univ. Sussex (UK); Caltech, JPL, NHS C, Univ. Colorado (USA). This
development has been supported by national funding agencies: CSA
(Canada); NAOC (China); CEA, CNES, CNRS (France); ASI (Italy); MCINN
(Spain); SNSB (Sweden); STFC, UKSA (UK); and NASA (USA).

\end{acknowledgements}

\bibliographystyle{aa}
\bibliography{ref}

\setcounter{figure}{0}
\renewcommand{\thefigure}{A\arabic{figure}}

\appendix
\section{SED tracks on SPIRE color-color diagram predicted by different templates}
\begin{figure*}
\centering
\includegraphics[width=0.65\linewidth]{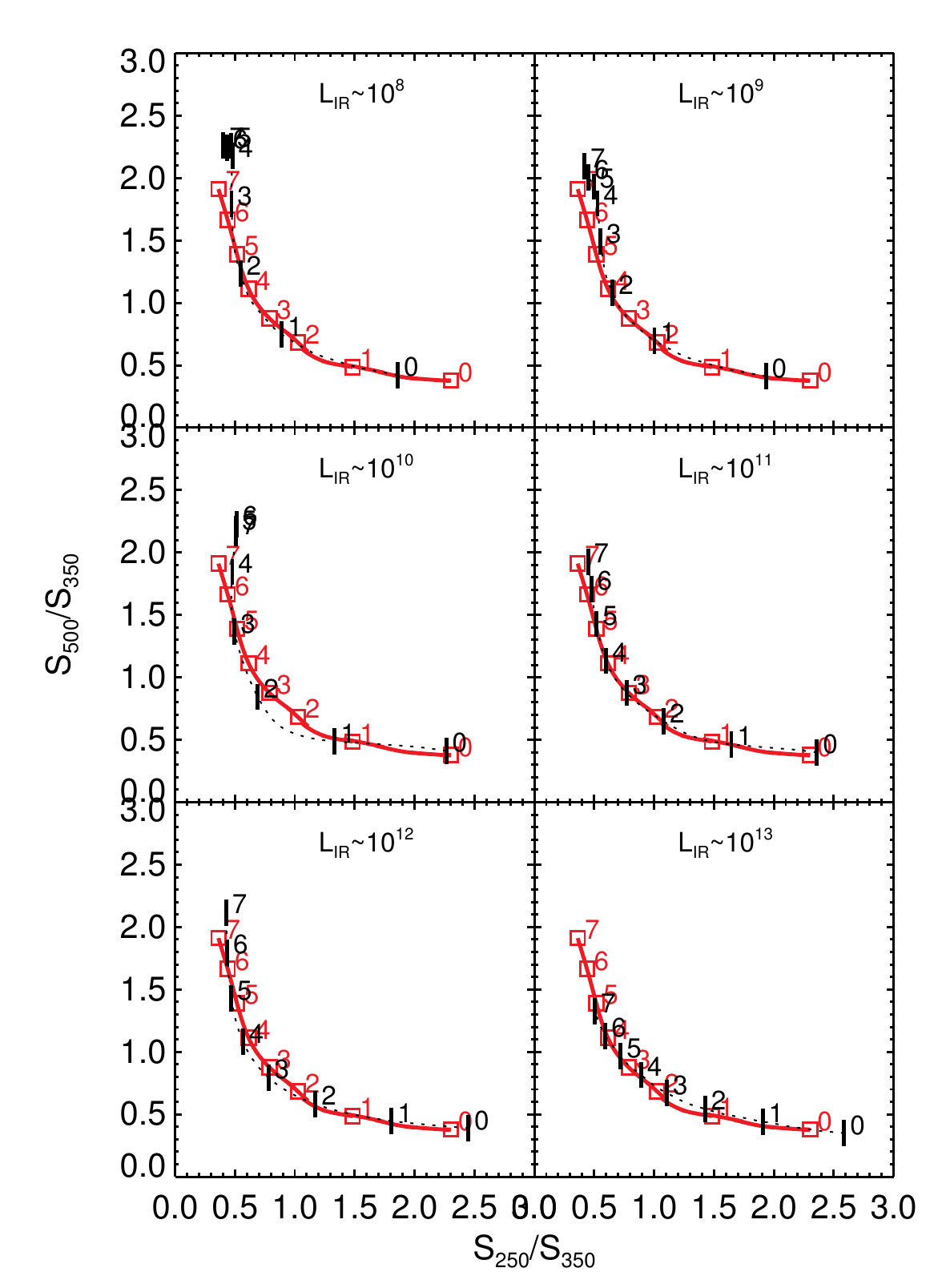}
\caption{SPIRE colors of the templates of CE01 (black dotted
  lines with vertical bars) with $L_{\rm IR}\sim 10^{8}$, $10^{9}$,
  $10^{10}$, $10^{11}$, $10^{12}$ , and $10^{13}$ $L_{\odot}$ . Red
  solid line with squares are the colors given by the MS
  templates of M12.}
\label{fig:sed_ce}
\end{figure*}

Figure \ref{fig:sed_ce} plots color-color relations for the templates
with $L_{\rm IR}\sim 10^{8}$, $10^{9}$, $10^{10}$, $10^{11}$,
$10^{12}$ , and $10^{13}$ $L_{\odot}$. $L_{\rm IR}$ $=$
$10^{11}L_{\odot}$ templates, and M12 MS templates produce very similar
SPIRE color-color relations.

\begin{figure*}
\centering
\includegraphics[width=0.65\linewidth]{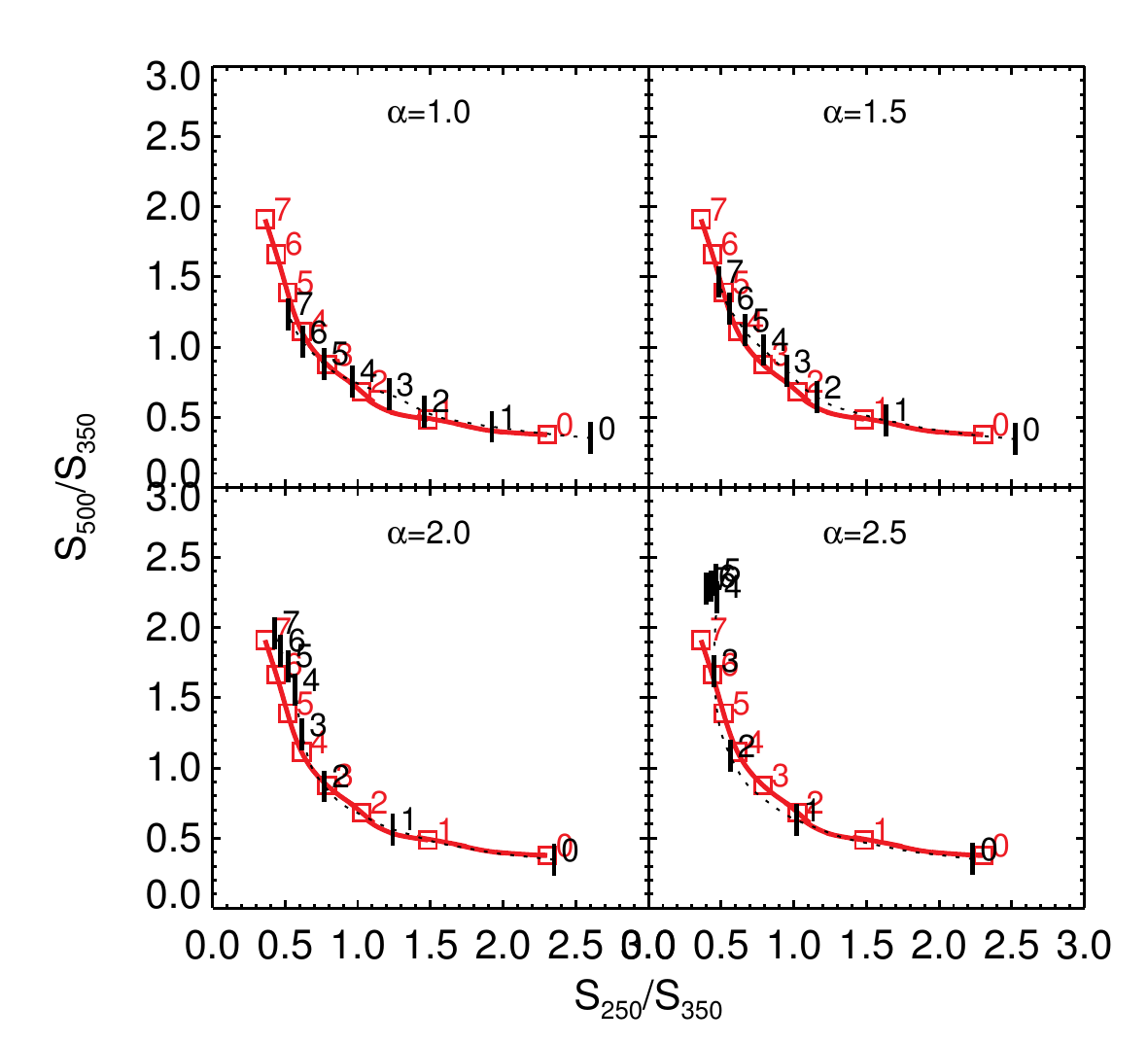}
\caption{SPIRE colors of the templates of DH02 templates
  (black dotted lines with vertical bars) with $\alpha$ = 1, 1.5, 2,
  and 2.5.  Red solid line with squares are the colors given by the
  MS templates of M12. }
\label{fig:sed_dh}
\end{figure*}

Figure \ref{fig:sed_dh} shows that the $\alpha\ =\ 2.0$ template
agrees well with M12 templates at the local universe. At $z\ =\ 2$,
$\alpha$ must reduce to $\sim$ $1.5$ to  agree with the evolution
trend suggested by M12 templates. The evolution of M12 MS templates is
similar to a variation of the value of $\alpha$ between $1.5$ to
$2.0$.

\begin{figure*}
\centering
\includegraphics[width=0.65\linewidth]{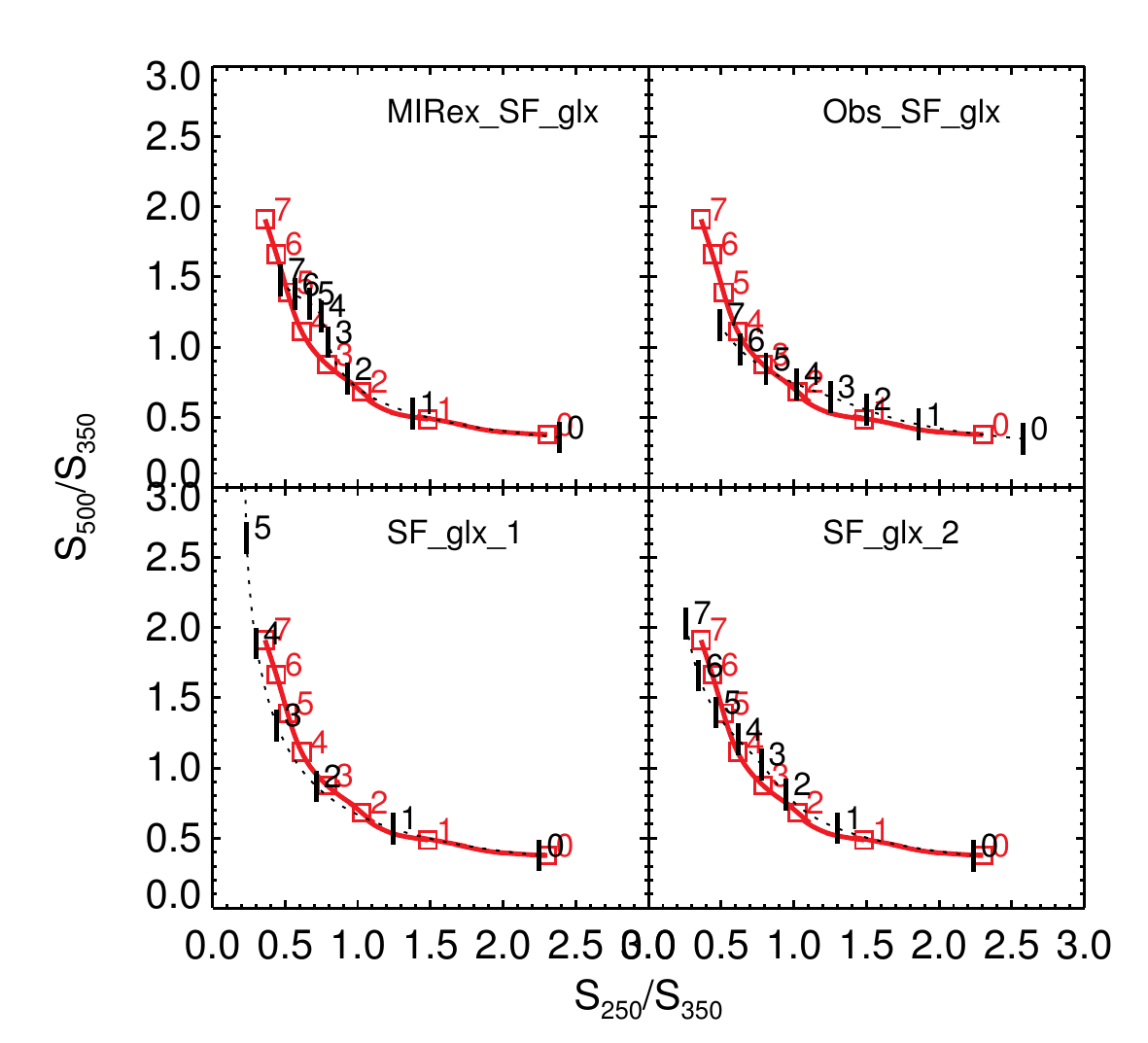}
\caption{SPIRE colors of the selected templates of B13
  templates (black dotted lines with vertical bars). Red solid line
  with squares are the colors given by the MS templates of
  M12.}
\label{fig:berta_sel}
\end{figure*}

Figure \ref{fig:berta_sel} shows the four templates selected from the
B13 libraries. The template SF\_glx\_2 (the template for
star-forming galaxies derived from COSMOS field) reproduces the SPIRE
colors very similar to the M12 MS templates at $z>4$. The SF\_glx\_1
template (star-forming galaxies derived from GOODS-S field) reproduces
much colder colors than the M12 templates. The MIRex\_SF\_glx
(MIR excess star-forming galaxies) and Obs\_SF\_glx (obscured
star-forming galaxies) templates have hotter SPIRE colors than the M12
MS templates.

\end{document}